\documentclass[aps,prb,reprint,amsmath,amssymb,showpacs,superscriptaddress,twocolumn]{revtex4-2}
\usepackage{graphicx}
\usepackage{xcolor}
\usepackage{subfigure}
\usepackage{comment}
\usepackage{braket}
\usepackage{amsmath}
\usepackage{commath}
\usepackage{mathtools}
\usepackage{chemformula}
\usepackage{natbib}
\usepackage{tabularx}
\usepackage{hyperref}
\usepackage{orcidlink}
\usepackage{booktabs}
\begin{document}
\title{Deterministic Switching of Perpendicular Ferromagnets by Higher harmonics of Spin-orbit Torque in Noncentrosymmetric Weyl Semimetals}
\author{Naomi Fokkens\,\orcidlink{0009-0002-1443-3651}} 
\affiliation{Department of Physics,	University of Alabama at Birmingham, Birmingham, AL 35294, USA}	
\author{Fei Xue\,\orcidlink{0000-0002-1737-2332}} 
\affiliation{Department of Physics,	University of Alabama at Birmingham, Birmingham, AL 35294, USA}	
\date{\today}
\begin{abstract}
Field-free deterministic switching of perpendicular ferromagnets is a central challenge for spintronics applications, typically requiring explicit symmetry breaking. Here we show that deterministic switching can instead be achieved through higher angular harmonics of spin-orbit torques, even in systems that preserve in-plane mirror symmetries. Using a vector spherical harmonics expansion, we demonstrate that these higher-harmonic torque components naturally give rise to additional out-of-equator fixed points, enabling reliable magnetization reversal when their magnitude is comparable to conventional lowest-order torques. We illustrate this mechanism with first-principles calculations on the noncentrosymmetric Weyl ferromagnet PrAlGe, where the combination of
Weyl-node band topology and strong spin-orbit coupling produces sizable higher-harmonic torque components. Because the Fermi surface is small, the conventional lowest-order torques are relatively weak, allowing the higher-order harmonics to compete on equal footing and strongly reshape the magnetization dynamics.
The resulting spin dynamics confirm deterministic switching without additional symmetry breaking. Our results establish higher-harmonic spin-orbit torque as a key ingredient for understanding and controlling magnetization dynamics in topological and spintronic materials.

\end{abstract}
\maketitle

\section{Introduction}
Realizing electric field control of magnetization without an external magnetic field is a central goal in spintronics applications such as magnetic random access memories and neuromorphic computing ~\cite{Manchon2019review,Grollier2020,Shao2021,Hoffmann2022}. Spin-orbit torque (SOT) has emerged as a promising mechanism for such control~\cite{miron2011perpendicular,liu2012spin}, requiring the coexistence of spin-orbit coupling, magnetization, and broken inversion symmetry. In prototypical heavy-metal/ferromagnet bilayers, an in-plane electric field can induce a spin current in the heavy metal layer that exerts a torque on the adjacent ferromagnet. Because the nonequilibrium spin accumulation can align with the magnetization for certain orientations, the torque vanishes in those cases, and its form is therefore strongly constrained by the underlying crystal symmetry.

In conventional bilayer systems with continuous rotational symmetry, the spin-orbit torque vanishes when the magnetization lies in-plane. As a result, ferromagnets with perpendicular magnetic anisotropy (PMA) exhibit non-deterministic switching: once the electric field is removed, the magnetization may relax either back to the original or into the reversed easy axis. This limitation is critical, since PMA is central for high-density, thermally stable spintronic devices~\cite{Worledge2011,Dieny2017,Shao2021}. The common strategy to achieve deterministic switching is to break in-plane mirror symmetry, either externally by applying an in-plane magnetic field~\cite{miron2011perpendicular,liu2012current} or intrinsically by utilizing reduced-symmetry materials~\cite{MacNeill2016,Xue2020SOT,Kao2022,Wang2022Cascadable,Sarkar2024}.

In this work, we demonstrate that the deterministic switching of ferromagnets with perpendicular magnetic anisotropy can be realized even \textit{without} breaking in-plane mirror symmetry by  incorporating higher harmonics of the linear-response spin-orbit torque (SOT). Throughout this work, we use “higher harmonics” (or equivalently “higher-order angular terms”) to refer to higher-order terms in the magnetization-angle dependence of the SOT, rather than to nonlinear
response in the applied electric field.
These symmetry-allowed contributions generate additional fixed points away from the equator, in contrast to the conventional lowest-order torques that enforce only in-plane fixed points. Whether the magnetization relaxes to these out-of-equator states depends on the relative strength of higher- versus lower-harmonic SOT terms. Using Landau-Lifshitz-Gilbert simulations of a minimal toy model containing both contributions, we construct a dynamical phase diagram that identifies the conditions under which deterministic switching occurs.

Higher-harmonic SOTs naturally arise from strong spin-orbit coupling effects beyond the linear perturbative regime~\cite{Mahfouzi2020,go2020theory}. To illustrate this mechanism in a real material, we consider the noncentrosymmetric Weyl ferromagnet  PrAlGe~\cite{Chang2018,Meng2019,Sanchez2020, Destraz2020,Liu2021,Yang2022,Forslund2025}, which exhibits both strong spin-orbit coupling and $C_{4z}$ rotational symmetry. First-principles calculations reveal that higher-harmonic SOT terms at the Fermi level are comparable in magnitude to conventional ones, enabling the deterministic switching without any additional symmetry breaking.

This paper is organized as follows. In Sec.~\ref{sec:symm},  we analyze the symmetry constraints on spin-orbit torque and express the allowed forms using vector spherical harmonics.  In Sec.~\ref{sec:toymodel}, we construct a minimal toy model combining lower- and higher-harmonic torques, and use Landau-Lifshitz-Gilbert simulations to map out the dynamical phase diagram, identifying the conditions for deterministic switching. Sec.~\ref{sec:firstprinc} presents first-principles calculations of spin-orbit torques in PrAlGe, followed in Sec.~\ref{sec:dynamics} by an analysis of the resulting spin dynamics and the role of higher-harmonic contributions in producing both deterministic reversal and sustained precession. Finally, Sec.~\ref{sec:discussion} summarizes the conclusions, discusses experimental relevance, and outlines future directions for exploring higher-harmonic torque effects.

\section{Symmetry-Allowed Spin-Orbit Torque} \label{sec:symm}
\begin{figure*}[htbp]
    \centering
    \includegraphics[width=0.9\textwidth]{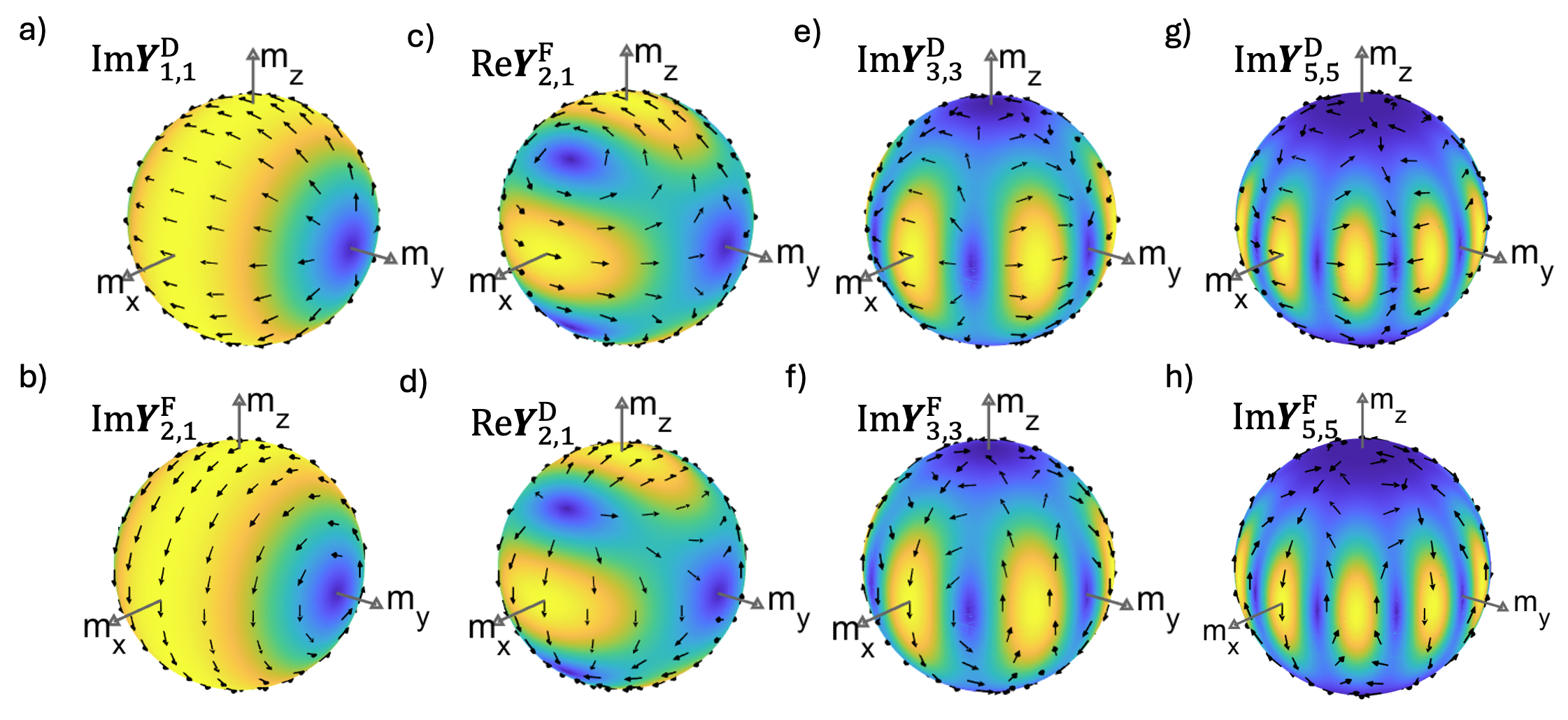} 
    \caption{Angular dependence of representative spin-orbit torque components for an applied electric field along $\hat{\mathbf{x}}$. 
Arrows indicate the torque direction on the unit sphere of magnetization $\hat{\mathbf{m}}$, while the color scale denotes torque magnitude. 
Panels (a) and (b) show the conventional lowest-order dampinglike 
($\mathrm{Im}\,\mathbf{Y}^{\mathrm{D}}_{1,1}\propto \hat{\mathbf{m}}\times(\hat{\mathbf{y}}\times\hat{\mathbf{m}})$) 
and fieldlike ($\mathrm{Im}\mathbf{Y}^{\mathrm{F}}_{1,1}\propto \hat{\mathbf{y}}\times\hat{\mathbf{m}}$) torques, 
which stabilize only equatorial fixed points. 
Panels (c)-(h) illustrate selected higher-harmonic symmetry-allowed torques, 
including $\mathrm{Re}\,\mathbf{Y}^{\mathrm{F}}_{2,1}$, $\mathrm{Re}\,\mathbf{Y}^{\mathrm{D}}_{2,1}$, $\mathrm{Im}\,\mathbf{Y}^{\mathrm{D}}_{3,3}$, 
$\mathrm{Im}\,\mathbf{Y}^{\mathrm{F}}_{3,3}$,
$\mathrm{Im}\,\mathbf{Y}^{\mathrm{D}}_{5,5}$, and
$\mathrm{Im}\,\mathbf{Y}^{\mathrm{F}}_{5,5}$,
Regions in blue correspond to vanishing torque, indicating fixed points. 
These higher-harmonic contributions introduce off-equator fixed points that play a crucial role in enabling deterministic switching when comparable in strength to the conventional terms.}
    \label{fig:AllowedTorque}  
\end{figure*}

In rotationally symmetric systems with broken inversion symmetry (with the symmetry-breaking direction denoted $\hat{\mathbf{n}}$), the lowest-order spin-orbit torques take two characteristic forms~\cite{Garello2013,Belashchenko2020,Xue2023}: $\hat{\mathbf{p}} \times \hat{\mathbf{m}}$ and $\hat{\mathbf{m}} \times (\hat{\mathbf{p}} \times \hat{\mathbf{m}})$, where $\hat{\mathbf{p}} = \hat{\mathbf{n}} \times \hat{\mathbf{E}}$. As illustrated in Fig.~\ref{fig:AllowedTorque}~(a,b) for $\hat{\mathbf{n}} = \hat{\mathbf{z}}$, these correspond to the conventional fieldlike and dampinglike torques. Their angular dependence resembles that of a static magnetic field and a Landau-Lifshitz damping term, respectively. Together, they stabilize an in-plane fixed point on the equator, $\hat{\mathbf{m}} = \hat{\mathbf{p}}$, where the total torque vanishes.

For field-free deterministic switching of perpendicular ferromagnets, the torque must drive the magnetization from its initial easy axis to the opposite axis under an applied electric field. This requires the existence of a stable fixed point \textit{off} the equator in the opposite hemisphere, so that once the field is removed, the magnetization relaxes into the reversed easy-axis state. Reversing the polarity of the electric field then drives the magnetization back, enabling fully bidirectional control solely through the applied field.

In systems with full rotational symmetry, however, all lowest-order spin-orbit torques  vanish when $\hat{\mathbf{m}}$ lies in‐plane, preventing the formation of off-equator fixed points.  One approach is to break an in-plane mirror symmetry so that $\hat{\mathbf{p}}$ acquires an out-of-plane component~\cite{Xue2020SOT,MacNeill2016}. Yet even in materials such as \ch{Fe3GeTe2}, which intrinsically breaks an in-plane mirror symmetry, lowest-order spin-orbit torques still stabilize only equatorial fixed points~\cite{Brataas2019}. As shown in Ref.~\cite{Xue2023}, off-equator fixed points appear only when higher-harmonic, magnetization-dependent torque terms are included. 
These observations motivate the central question of this work: whether higher-harmonic spin-orbit torques, without explicitly breaking any in-plane symmetry, can generate off-equator fixed points that enable deterministic switching of a perpendicular ferromagnet.

To answer this, we systematically derive the symmetry-allowed forms of spin-orbit torque, using a vector spherical harmonics expansion~\cite{Belashchenko2020,Xue2023,Fang_2025}. Unlike conventional Cartesian expansions, this formalism provides a complete orthonormal basis without introducing artificial cutoffs and naturally separates torques into fieldlike and dampinglike components. In addition, it offers a transparent way to identify and classify higher-harmonic contributions consistent with the crystal symmetry, making it particularly suited for the present study.

We focus on systems with $C_{4z}$ symmetry, with both $xz$ and $yz$ mirror planes. This choice isolates higher-harmonic torque effects: unlike $C_{3z}$ or other lower-symmetry groups, $C_{4z}$ does not break any in-plane mirror symmetry, and unlike continuous rotational symmetry, it does not forbid azimuthal angular dependence~\cite{Belashchenko2020}. Thus $C_{4z}$ is the minimal symmetry setting where higher-harmonic torques can be unambiguously disentangled from symmetry breaking.

The spin-orbit torque is described in linear response by $\mathcal{T}_i=\tau_{ij}E_j$ (or equivalently $\boldsymbol{\mathcal{T}}
=\boldsymbol{\tau}_{\mathbf{\hat{E}}}\mathbf{E}$), where spin-orbit torkance tensor can be expanded in vector spherical harmonics following the methodology of Ref.~\cite{Xue2023}, with system-specific details given in Appendix~\ref{appendix:SymmAnl}. Applying this expansion under the symmetry constraints for an electric field along $\hat{\mathbf{x}}$ yields the allowed higher-harmonic torque components, which separate naturally into time-reversal even and odd contributions:

\begin{equation}
\begin{aligned}
\boldsymbol{\tau}^{\text{even}}_{\hat{\mathbf{x}}} = \sum_{l,m}
[ &C^{\rm F}_{2l,4m\pm1}\operatorname{Re}\mathbf{Y}^{\rm F}_{2l,4m\pm1} \\
+&C^{\rm D}_{2l+1,4m\pm1}\operatorname{Im}\mathbf{Y}^{\rm D}_{2l+1,4m\pm1}],
\end{aligned}
\label{eq:tau_even}
\end{equation}
\begin{equation}
\begin{aligned}
\boldsymbol{\tau}^{\text{odd}} _{\hat{\mathbf{x}}} = \sum_{l,m} [ &C^{\rm D}_{2l,4m\pm1}\operatorname{Re}\mathbf{Y}^{\rm D}_{2l,4m\pm1} \\
+&C^{\rm F}_{2l+1,4m\pm1}\operatorname{Im}\mathbf{Y}^{\rm F}_{2l+1,4m\pm1}].
\end{aligned}
\label{eq:tau_odd}
\end{equation}

Equations~\ref{eq:tau_even} and~\ref{eq:tau_odd} consist of a symmetry-allowed combination of fieldlike and dampinglike terms, with coefficients $C$ specifying the weight of each contribution. For an electric field applied along $\hat{\mathbf{y}}$, the corresponding torque forms follow directly from Eqs.~\ref{eq:tau_even}-\ref{eq:tau_odd} by applying the $C_{4z}$ rotational symmetry. The explicit expressions are provided in Appendix~\ref{appendix:SymmAnl}.

The symmetry-allowed forms of spin-orbit torque are illustrated in Fig.~\ref{fig:AllowedTorque}. Panels (a) and (b) show the conventional lowest-order fieldlike and dampinglike terms, which enforce only equatorial fixed points. Panels (c)-(h) present representative examples of higher-harmonic torques permitted by $C_{4z}$ symmetry. In each plot, arrows indicate the torque direction and the color scale denotes magnitude, with blue regions corresponding to vanishing torque and thus potential fixed points. Because the vector spherical harmonics expansion includes contributions of arbitrarily high order, only a subset of symmetry-allowed terms relevant for the present analysis is shown here.
In the next section, we examine the dynamical consequences of selected higher-harmonic terms by constructing a minimal toy model, which highlights how these contributions modify fixed points and enable deterministic switching.

\begin{figure}[htbp]
    \centering
    \includegraphics[width=1\linewidth]{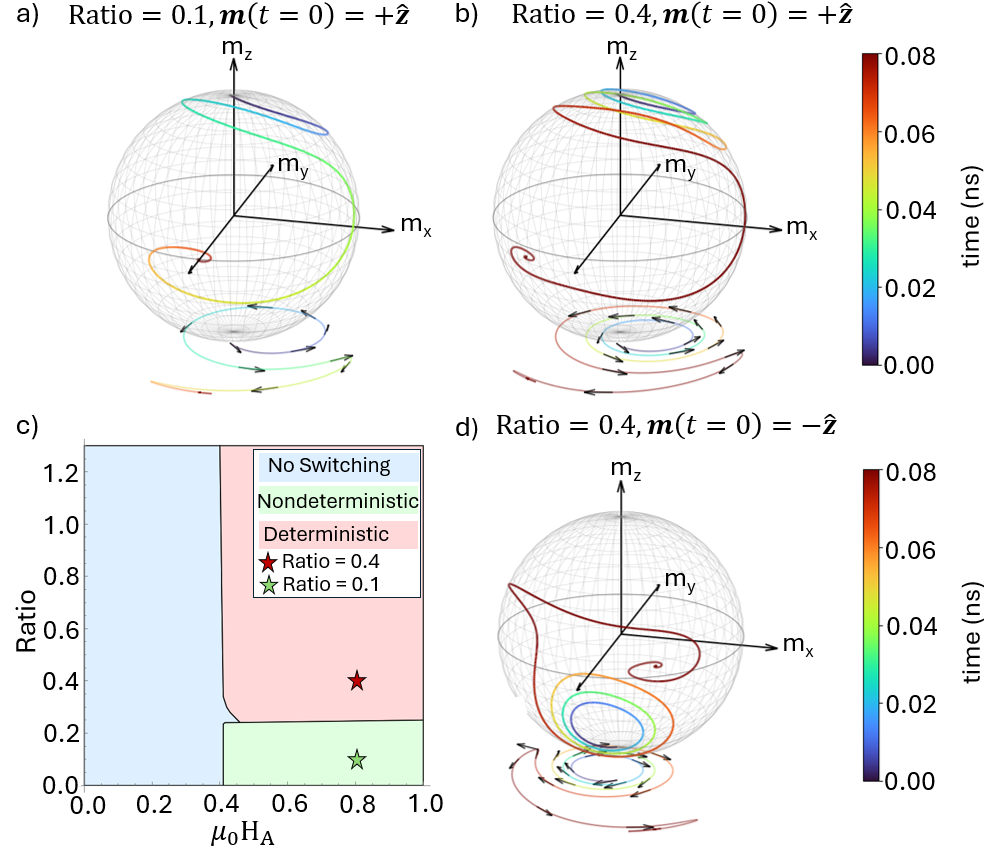} 
\caption{LLG simulations of magnetization dynamics with conventional lowest-order torques [Fig.~\ref{fig:AllowedTorque}(a,b)] and the selected higher-harmonic torque component [Fig.~\ref{fig:AllowedTorque}(f)]. Panels (a) and (b) show representative trajectories for ratios of higher- to lower-harmonic torque amplitudes of $0.1$ and $0.4$, respectively. For a small ratio (a), the magnetization is driven to the equatorial fixed point, resulting in nondeterministic switching. For a larger ratio (b) the trajectory crosses the equator and relaxed into a reversed state, demonstrating deterministic switching. Panel (c) presents the phase diagram of switching regimes—deterministic, nondeterministic, and no switching—as functions of the applied electric field $E$ relative to the anisotropy field $\mu_0H_{\rm A}$ and the ratio of higher-harmonic contributions. Panel (d) shows that starting from the south pole, the same electric-field direction drives the magnetization to the symmetry-related north-hemisphere fixed point, illustrating symmetry-preserving deterministic switching without reversing the field polarity.}
    \label{fig:2} 
\end{figure}

\section{Deterministic Switching in a Landau-Lifshitz-Gilbert Model with Higher-Harmonic Torques}\label{sec:toymodel}

Having established the symmetry-allowed forms of spin-orbit torque, we next examine their dynamical consequences using a minimal Landau-Lifshitz-Gilbert (LLG) model. Our goal is to identify the simplest combination of torque harmonics capable of producing deterministic magnetization reversal in a system that preserves global $C_{4z}$ and mirror symmetries. We focus on a minimal torque expansion

\begin{equation}
\boldsymbol{\tau} =  \text{Im}  \mathbf{Y}^{\rm D}_{1,1}+\text{Im}  \mathbf{Y}^{\rm F}_{1,1} + \text{ratio}*
\text{Im}  \mathbf{Y}^{\rm F}_{3,3}.
\label{eq:toymodel}
\end{equation}
The first two terms, $\rm{Im}\mathbf{Y}^{\rm D}_{1,1}$ and $\rm{Im}\mathbf{Y}^{\rm F}_{1,1}$, represent the conventional lowest-order dampinglike and fieldlike torques, which are expected to dominate magnetization dynamics. They generate fixed points confined to the $\mathbf{m}\parallel \hat{\mathbf{y}}$, as shown in Fig.~\ref{fig:AllowedTorque}~(a-b).  In this situation a dc electric field cannot uniquely steer the magnetization from $+\hat{\mathbf{z}}$ to a definite state in the opposite hemisphere. Depending sensitively on small perturbations, the magnetization either returns to the original side or relaxes to the opposite easy-axis (nondeterministic switching).

Introducing higher-order angular harmonics ($\ell > 1$) adds additional fixed points off the equator [Figs.~\ref{fig:AllowedTorque}(c-h)], but they appear symmetrically in both hemispheres due to the preserved mirror symmetries $\sigma_{xz}$ and $\sigma_{yz}$, leaving the dynamics degenerate.
The key question, then, is whether the combined torque field drives $\hat{\mathbf{m}}$ toward a same-hemisphere fixed point (no switching) or across the equator into its mirror-related counterpart (deterministic switching).

A crucial ingredient is the inclusion of a higher-order term with a different azimuthal structure, such as $\mathrm{Im}\mathbf{Y}^{\rm F}_{3,3}$ [Fig.~\ref{fig:AllowedTorque}(f)].
Unlike the conventional $m=1$ torques, which generate single-lobed angular patterns, this $m=3$ component redistributes the torque on the unit sphere so that the flow lines naturally cross the equator rather than remaining trapped within one hemisphere.
When combined with the $m=1$ torques, it modifies the global flow on the unit sphere: trajectories starting near $+\hat{\mathbf{z}}$ are guided toward a unique off-equator fixed point in the opposite hemisphere, while those starting near $-\hat{\mathbf{z}}$ relax toward its mirror-related partner. 
Importantly, the torque still vanishes at $\hat{\mathbf{m}}=\pm\hat{\mathbf{y}}$ by mirror symmetry, so deterministic switching only occurs when the initial magnetization lies away from this high-symmetry direction.

The magnetization dynamics in our toy model are governed by the Landau-Lifshitz-Gilbert (LLG) equation~\cite{SLONCZEWSKI1996}:
\begin{equation}
\frac{d\hat{\mathbf{m}}}{dt} - \alpha \hat{\mathbf{m}} \times \frac{d\hat{\mathbf{m}}}{dt} 
= -\gamma \mu_0 H_{\rm A} (\hat{\mathbf{m}} \times \hat{\mathbf{z}})(\hat{\mathbf{m}} \cdot \hat{\mathbf{z}}) + \boldsymbol{\mathcal{T}}
\label{eq:LLG}
\end{equation}
where $\hat{\mathbf{m}}$ denotes the magnetization direction, $\gamma$ is the gyromagnetic ratio, $\hat{\mathbf{z}}$ specifies the easy axis and $\boldsymbol{\mathcal{T}}$ represents the spin-orbit torque. In the simulations, the damping parameter is fixed at $\alpha = 0.01$, and the anisotropy field is set to $\mu_0H_{\rm A} = 2$~T, oriented along the perpendicular $\hat{\mathbf{z}}$ direction.

Figure~\ref{fig:2} summarizes the numerical results of the toy model. Panel (a) shows the case of a relatively small ratio ($0.1$) of higher-harmonic torque contributions to the conventional lowest-order terms. In this regime, the magnetization is driven only to the symmetry-enforced fixed point at the equator, resulting in nondeterministic switching.
In contrast, panel (b) illustrates the case of a larger ratio ($0.4$) of higher-harmonic to lowest-order terms. Here, the magnetization crosses the equator and settles into a new stable off-equator fixed point created by the higher-harmonic torques. When the electric field is removed, the magnetization reliably relaxes into the reversed easy-axis direction, thereby realizing deterministic switching.
Panel (c) presents the numerical phase diagram mapping the switching regimes—deterministic, nondeterministic, and no switching—as functions of the applied electric field strength ($E$) relative to the anisotropy field ($\mu_0 H_{\rm A}$) and the relative weight (“ratio”) of higher-harmonic torque terms.

Importantly, this deterministic switching mechanism is not unique to the specific $\mathrm{Im}\mathbf{Y}^{\mathrm{F}}_{3,3}$ term used in the toy model. Other higher-order harmonics with different azimuthal structures can generate similar behavior: for instance, the DFT-derived torque coefficients in PrAlGe presented later in Sec.~\ref{sec:dynamics} show that $\mathrm{Im}\mathbf{Y}^{\mathrm{D}}_{5,5}$ is sufficiently large to induce deterministic switching under realistic conditions, while additional components can lead to oscillatory dynamics (see Appendix~\ref{appendix:LLGPhaseDiagrams}).
This generality underscores that deterministic switching does not rely on a single harmonic, but on the interplay between conventional and higher-harmonic torque symmetries that promote magnetization flow across the equator.

Finally, this “symmetry-preserving” switching mechanism differs fundamentally from the conventional symmetry-breaking case.
In broken-mirror systems, reversing the magnetization between north and south poles requires reversing the electric-field direction, as each field polarity favors only one hemisphere (governed by mirror $\sigma_{yz}$ symmetry). In that case, explicit mirror-symmetry breaking allows the applied electric field to generate an effective north–south energetic bias,
lifting the equivalence of the two hemispheres for a given field
polarity and thereby selecting a unique switching direction.

Here, by contrast, the same field direction drives reversal from either pole [Fig.~\ref{fig:2}(d)]: an initial state and its mirror partner (related by $\sigma_{xz}$ mirror symmetry: $\theta\to\pi-\theta$, $\phi\to\pi-\phi$) evolve toward final states that are themselves mirror related. 
Reversing the field still reverses the switching trajectory, but now the magnetization is driven toward a different fix point related by the $\sigma_{yz}$ mirror. 
Importantly,
this switching does not rely on an electric-field-induced north–south
energetic bias between the $+m_z$ and $-m_z$ hemispheres. Instead, 
the two hemispheres remain symmetry related, and deterministic switching 
is selected by which symmetry-related off-equator fixed point the initial 
magnetization flows toward under the torque field.
Thus deterministic switching arises not from breaking symmetry due to 
applied electric field, but from the internal angular structure of 
the spin–orbit torque itself (see Appendix~\ref{appendix:LLG} for further discussion).

\section{First-Principles Electronic Structure and Wannierization of PrAlGe} \label{sec:firstprinc}

\begin{figure*}[htbp]
    \centering
    \includegraphics[width=0.9\textwidth]{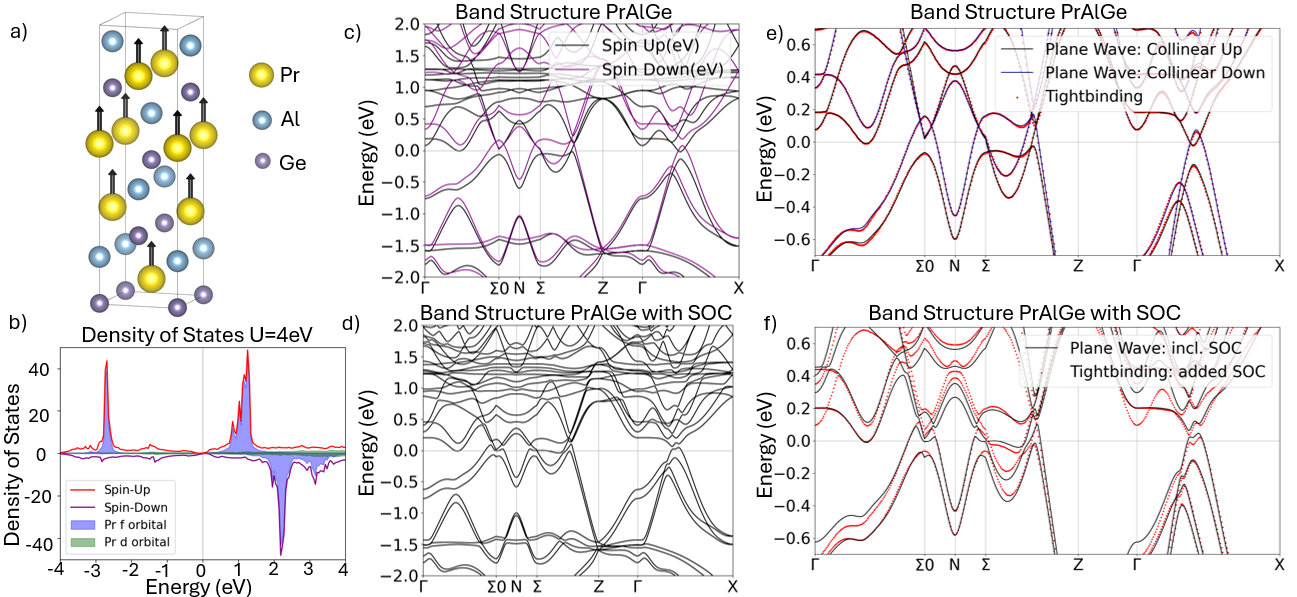} 
\caption{
(a) Crystal structure of PrAlGe with out-of-plane Pr magnetic moments. 
(b) Spin-resolved density of states calculated with Hubbard $U=4$ eV, highlighting the Pr $f$ orbitals. 
(c,d) Electronic band structures of PrAlGe from plane-wave DFT: (c) collinear calculation without spin-orbit coupling (SOC) and (d) noncollinear calculation including SOC. 
(e) Comparison of symmetrized Wannier-interpolated bands with the collinear plane-wave results (c). 
(f) Comparison of Wannier-interpolated bands with SOC added at the atomic level to the collinear model, benchmarked against the noncollinear plane-wave results (d). 
Panels (e,f) are shown in a reduced energy range near the Fermi level to highlight the quality of the Wannier fitting. 
}
    \label{fig:3}  
\end{figure*}

Motivated by the symmetry analysis and toy model results, we now turn to PrAlGe, a noncentrosymmetric ferromagnetic Weyl semimetal from the \textit{R}AlGe family that provides an ideal platform to test the role of higher-harmonic spin-orbit torques. PrAlGe possesses $C_{4z}$ rotational symmetry, simultaneously breaks time-reversal and inversion symmetries (stabilizing Weyl nodes), and exhibits strong spin-orbit coupling, making it a promising candidate for spintronic applications~\cite{Chang2018,Sanchez2020, Destraz2020}. These properties suggest that higher-harmonic torque contributions may be sizable, and if comparable to the lowest-order terms, could enable deterministic switching consistent with the toy model predictions.

To evaluate spin-orbit torques in PrAlGe, we first construct a reliable electronic structure model based on density functional theory (DFT). PrAlGe crystallizes in the noncentrosymmetric tetragonal space group $I4_1md$ (No.~109), which preserves $C_{4z}$ rotational symmetry. Calculations are performed within the GGA+$U$~\cite{GGA+U} framework using VASP~\cite{VASP} ~\cite{Blochl1994} ~\cite{Kresse1999} (computational parameters are given in Appendix~\ref{appendix:FPC}). A Hubbard $U=4$ eV is applied to the Pr $f$-orbitals, consistent with prior studies~\cite{Chang2018,Sanchez2020}, which effectively shifts the localized $f$ states away from the Fermi level [Fig.~\ref{fig:3}~(b)]. This ensures that the low-energy electronic structure is dominated by itinerant $spd$ states, providing a reasonable basis for transport and torque modeling. Both collinear and noncollinear calculations are carried out, and the resulting crystal structure, density of states, and band dispersions [Figs.~\ref{fig:3}~(a-d)] are consistent with earlier studies of the \textit{R}AlGe family~\cite{Chang2018,Sanchez2020}. In particular, the band structure shows the expected semimetallic character with linearly dispersing crossings near the Fermi level that evolve into Weyl points when spin-orbit coupling and ferromagnetism are included, in agreement with prior reports.

From the collinear DFT results, we construct a tight-binding model using Wannier90~\cite{Wannier90} and symmetrize it with WannSymm~\cite{Zhi2022WannSymm} to enforce the crystal symmetry. The Wannier-interpolated bands reproduce the collinear DFT dispersion near the Fermi level [Fig.~\ref{fig:3}(e)]. Spin-orbit coupling is then included as an on-site $\alpha \mathbf{L}\cdot\mathbf{S}$ term~\cite{Kurita2020,Tartaglia2020} and benchmarked against the noncollinear DFT results [Fig.~\ref{fig:3}(f)]. This procedure is advantageous because Wannierization of the collinear Hamiltonian is more stable and keeps the exchange term purely spinlike, enabling straightforward rotation of the magnetization in torque calculations. At the same time, the atomic SOC approach reproduces the essential band features of the noncollinear calculation, providing a reliable model for analyzing spin-orbit torque and its angle dependence. Additionally, this explicit on-site SOC form further allows controlled SOC scaling tests of the torkance (detailed in Appendix~\ref{appendix:tuneSOC}).

This Wannier-based model captures the essential electronic and spin-dependent properties of PrAlGe. In particular, Appendix~\ref{appendix:WP} confirms that the Weyl points obtained from the noncollinear VASP calculations are faithfully reproduced in the symmetrized tight-binding model with atomic SOC, ensuring that no topological features are lost in the fitting procedure. This agreement provides a robust foundation for calculating spin-orbit torques and quantifying the magnitude of higher-harmonic torque components.

\section{Spin Dynamics and Deterministic Switching in PrAlGe}\label{sec:dynamics}

With the validated Wannier tight-binding Hamiltonian in hand, we now turn to the calculation of spin-orbit torques and the resulting magnetization dynamics. Using the symmetry framework established in Sec.~\ref{sec:symm}, the torque is expanded in a vector spherical harmonics basis, allowing us to decompose the contributions into lowest-order (conventional) and higher-order angular components. We then analyze how these torques govern magnetization dynamics in PrAlGe by solving the Landau-Lifshitz-Gilbert (LLG) equation, directly connecting the symmetry-allowed torque terms to the possibility of deterministic switching.

The standard Kubo formula~\cite{Freimuth2014,Mahfouzi2018,Belashchenko2019,Xue2023} is used to evaluate the time-reversal even and odd components of the spin-orbit torkance in PrAlGe, based on the \textit{ab initio} tight-binding Hamiltonian constructed in Sec.~\ref{sec:firstprinc}:

\begin{equation}
		\label{eq:eventorkance}
		\tau^{\rm even}_{ij}=2e\sum_{\substack{\mathbf{k},n,\\
				m\neq n}} f_{n\bf{k}} \frac{\text{Im}\bra{\psi_{n\bf{k}}}  \frac{\partial H_{\bf k}}{\partial k_j}\ket{\psi_{m\bf{k}}}\bra{\psi_{m\bf{k}}}  \mathcal{T}_i\ket{\psi_{n\bf{k}}}}{(E_{m\mathbf{k}}-E_{n\mathbf{k}})^2+\eta^2},
	\end{equation}

    \begin{equation}
    \label{oddtorkance}
\tau^{\text{odd}}_{ij} = \frac{e}{\pi} 
\sum_{\mathbf{k},n,m} 
\frac{\eta^2 \, \mathrm{Re} \left[ 
\langle \psi_{n\mathbf{k}} | \mathcal{T}_i | \psi_{m\mathbf{k}} \rangle 
\langle \psi_{m\mathbf{k}} | \frac{\partial H_{\bf k}}{\partial k_j} | \psi_{n\mathbf{k}} \rangle 
\right]}
{\left[ ( \mu -E_{n\mathbf{k}} )^2 + \eta^2 \right]
 \left[ ( \mu - E_{m\mathbf{k}} )^2 + \eta^2 \right]} .
 \end{equation}

Here $\ket{\psi_{n\mathbf{k}}}$\ are the Bloch eigenstates of the Hamiltonian $H_{\mathbf{k}}$, $E_{n\mathbf{k}}$ are the corresponding band
energies, and $f_{n\mathbf{k}}$ is the Fermi-Dirac function evaluated at the chemical potential $\mu$. The torque operator is defined as
$\boldsymbol{\mathcal{T}} = -\frac{\mathrm{i}}{\hbar}\,[\boldsymbol{\Delta}\cdot \hat{\mathbf{S}}, \hat{\mathbf{S}}],$
where $\hat{\mathbf{S}}$ is the spin operator and $\boldsymbol{\Delta}$ is the time-reversal odd exchange-correlation field that couples to spin. 
In practice, $\boldsymbol{\Delta}$ is extracted from the spin-dependent (TR-odd) part of the collinear Wannier Hamiltonian (exchange splitting) and is rotated rigidly with $\hat{\mathbf m}$ in the rotated-magnetization calculations.
The indices $i,j$ denote the directions of torque and  the applied electric field, respectively. The parameter
$\eta$ is the Lorentzian broadening used to model disorder and finite-temperature smearing (here we use $\eta = 25$ meV, $k_BT=1.6$ meV).
Additional numerical details of implementing these calculation are provided in Appendix~\ref{appendix:FPC}.

\begin{figure}[htbp]
    \centering
    \includegraphics[width=1\linewidth]{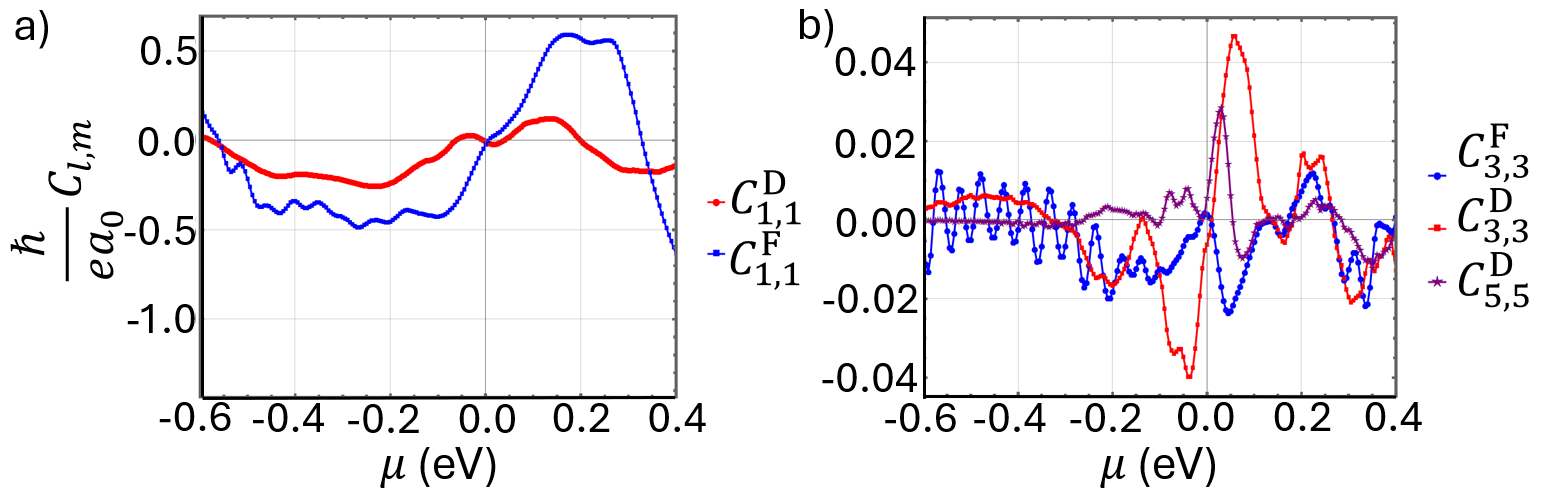} 
    \caption{Dependence of selected vector spherical harmonics coefficients on
    chemical potential $\mu$. 
    (a) Conventional lowest-order terms $C^{\rm D}_{1,1}$ and
    $C^{\rm F}_{1,1}$. 
    (b) Dominant higher-order terms $C^{\rm F}_{3,3}$,
    $C^{\rm D}_{3,3}$, and $C^{\rm D}_{5,5}$.}
    \label{fig:fxnMu}  
\end{figure}

We then project the \emph{ab initio} torques onto the symmetry-allowed vector spherical harmonics forms of Eqs.~\ref{eq:tau_even} and ~\ref{eq:tau_odd}. As expected, the extracted coefficients respect the 
$C_{4z}$ symmetry of PrAlGe, with only symmetry-allowed terms remaining. To assess the relative importance of different torque harmonics, it is useful to examine how the fitted coefficients $C_{\ell,m}(\mu)$ evolve as the chemical potential is varied. In any metallic system the spin-orbit torkance is an energy-dependent quantity, and scanning $\mu$ provides a systematic way to identify regimes where the higher-order components become comparable to the conventional lowest-order terms. The resulting trends are shown in Fig.~\ref{fig:fxnMu}. 
Panel~(a) illustrates that the conventional lowest-order terms  $C^{\rm D}_{1,1}$ and $C^{\rm F}_{1,1}$ are suppressed near the Fermi level, while panel~(b) shows that several higher-harmonic components—
notably the dampinglike $C^{\rm D}_{5,5}$ and, to a lesser extent, the fieldlike $C^{\rm F}_{3,3}$ and dampinglike $C^{\rm D}_{3,3}$—become  comparable in magnitude over the range $\mu \approx 0$-$0.1$ eV.

Importantly, the fitted coefficients at $\mu = 0.02$ eV
(Table~\ref{table:Fitting}) reveal that higher-harmonic terms make
contributions comparable in magnitude to the conventional components.
We see that the higher-order harmonic
emphasized in the toy model, $\mathrm{Im}\,\mathbf{Y}^{\rm F}_{3,3}$,
is present but relatively small. Instead, the dampinglike $\mathrm{Im}\,\mathbf{Y}^{\rm D}_{5,5}$ term is the dominant
higher-order contribution. As shown in the full torque field (Fig.~\ref{fig:TorqueSphereLLG}) and confirmed by the phase-diagram analysis in Appendix~\ref{appendix:LLGPhaseDiagrams}, a sufficiently large
$\mathrm{Im}\,Y^{\rm D}_{5,5}$ produces the similar off-equator fixed-point
structure and deterministic-switching behavior identified in the toy
model. This establishes $\mathrm{Im}\,Y^{\rm D}_{5,5}$ as the key
higher-order component governing the switching dynamics in PrAlGe.

\begin{table}
\centering
\begin{tabular}{| c | c | c | c | c | c | c |}
\hline
$C^{\rm D}_{1,1}$ &
$C^{\rm D}_{3,1}$ &
$C^{\rm D}_{3,3}$ &
$C^{\rm D}_{5,5}$ &
$C^{\rm D}_{2,1}$ &
$C^{\rm F}_{1,1}$ &
$C^{\rm F}_{3,3}$ 
 \\ \hline
-0.022 & 0.022 & 0.012 & 0.024 & 0.023 & 0.027 &  -0.008 
 \\ \hline
\end{tabular}
\caption{ The table shows the coefficients of the fitting for significant spin-orbit torque with units $e a_0/\hbar$ at $\mu=0.02$~eV. The first four are time-reversal even components while the last three are time-reversal odd components. The coefficients less than $0.01$ are dropped here for brevity.}
\label{table:Fitting}
\end{table}

\begin{figure}[htbp]
    \centering
    \includegraphics[width=1\linewidth]{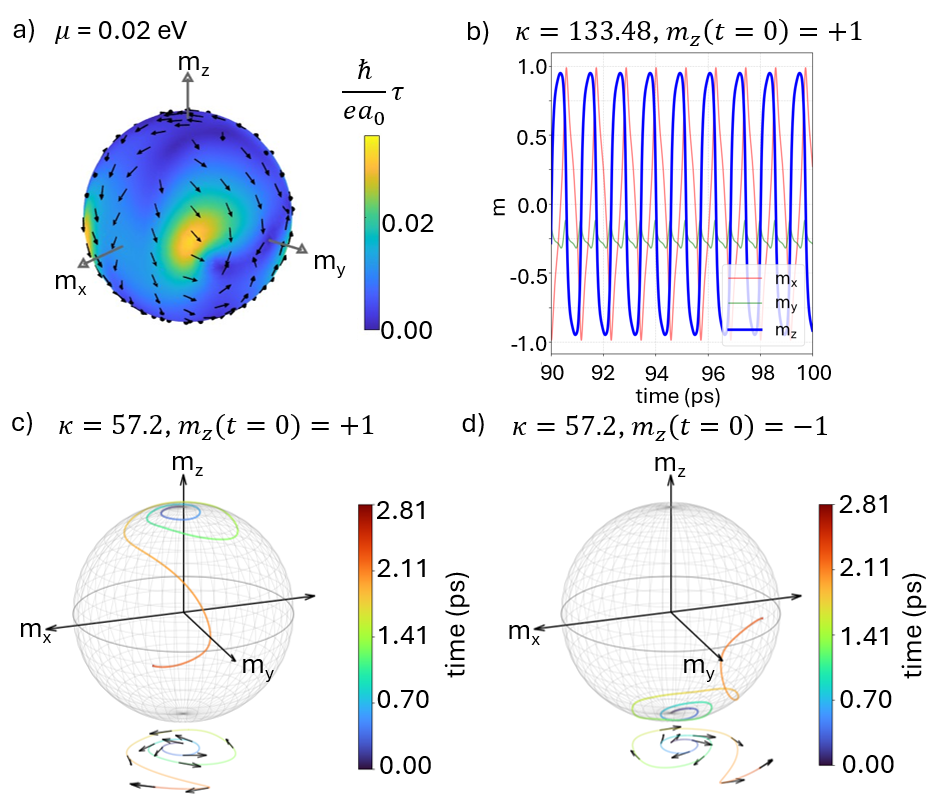} 
    \caption{Angular dependence of spin-orbit torque in PrAlGe at $\mu=0.02$ eV. 
    (a) spin-orbit torque in units $\tfrac{\hbar}{ea_0}$, evaluated for an applied electric field along $\hat{\mathbf{x}}$. $a_0$ is the atomic Bohr radius.
    The color scale indicates the torque magnitude on the magnetization unit sphere, while arrows denote the torque direction.
    (c,d) Magnetization trajectories obtained by solving the LLG equation
    with the \emph{ab initio} torque of panel (a) for
    $E \parallel \hat{\mathbf{x}}$, starting from initial states in the (c) northern
    and (d) southern hemispheres at the same electric field. Trajectories
    are drawn on the unit sphere; color encodes time.
    (b) For a larger electric field than in (c,d), the magnetization no
    longer relaxes to the off-equator fixed points but instead exhibits
    sustained precessional oscillations. The oscillation frequency is
    tunable with electric-field strength; for $\kappa = 133.48$ the frequency is
    approximately $0.85$ THz. Here $\kappa \equiv eEa_0/(\hbar\gamma \mu_0H_A)$, 
    is the dimensionless drive amplitude used in the LLG simulations (see main text).
    }
    
\label{fig:TorqueSphereLLG}  
\end{figure}

The calculated even and odd spin-orbit torques are combined and plotted on the magnetization unit sphere in Fig.~\ref{fig:TorqueSphereLLG}(a), revealing the full angular dependence of the SOT in PrAlGe. In contrast to the conventional lowest-order fieldlike and dampinglike forms in Fig.~\ref{fig:AllowedTorque}(a,b), the resulting spin-orbit torque exhibits pronounced higher-order angular structures with off-equator fixed points. The overall magnitude of the torque is relatively small compared to conventional bilayer system~\cite{Freimuth2014,Xue2020SOT}, consistent with the reduced density of states near the Fermi level in this Weyl semimetal. 
But its angular profile is qualitatively different: the higher-order harmonics
identified above—most notably the dampinglike
$\mathrm{Im}\,Y^{\rm D}_{5,5}$ term—substantially reshape the torque field. As shown below, this modified phase portrait is precisely what enables deterministic switching in PrAlGe.
\begin{figure}[htbp]
    \centering
    \includegraphics[width=0.5\textwidth]{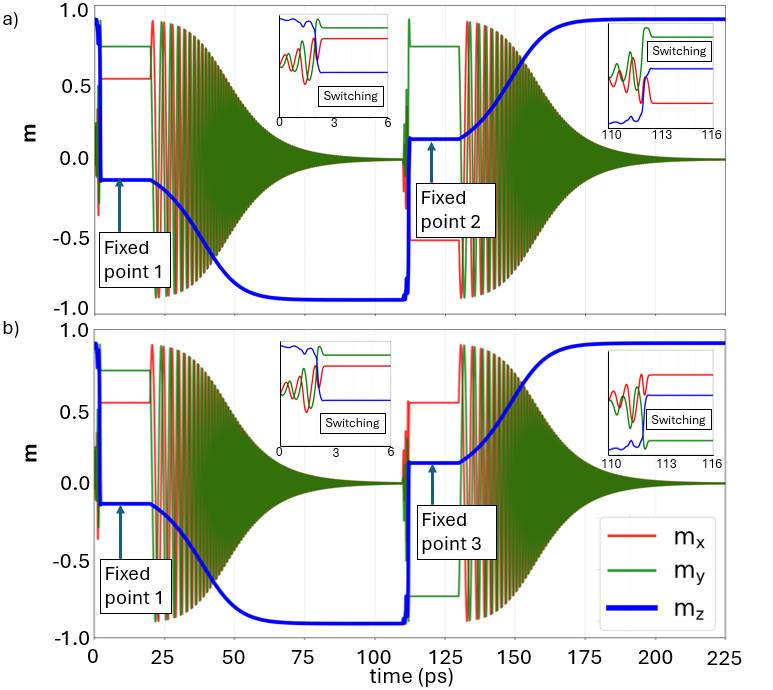} 
\caption{Reversible switching protocol using two electric-field pulses. Time evolution of the magnetization components $(m_x,m_y,m_z)$ under successive pulses of $\mathbf{E} \parallel \hat{\mathbf{x}}$, with relaxation intervals in between. In both panels the system starts near $+m_z$, is driven to an off-equator fixed point (labeled “Fixed point 1”), and relaxes to
$-m_z$ after the first pulse is turned off. In panel (a), a second pulse with the same polarity then drives the magnetization to a different off-equator fixed point in the northern hemisphere (“Fixed point 2”), from which it relaxes back to $+m_z$ when the field is removed.
In panel (b), the second pulse has the opposite polarity and accesses a third mirror-related fixed point in the northern hemisphere (“Fixed point 3”). Insets show the short-time switching dynamics to each fixed point. In both cases the protocol yields deterministic $+m_z \leftrightarrow -m_z$ switching, but through different fixed points.}
    \label{fig:EfieldOff} 
\end{figure}

To assess the dynamics, we solve the Landau-Lifshitz-Gilbert equation (Eq.~\ref{eq:LLG}) using the fitted torque coefficients and the DFT-derived uniaxial anisotropy field for PrAlGe, $\mu_0 H_{\rm A}=47.9\ \mathrm{T}$ (consistent with prior estimates~\cite{Skomski2009}), and a damping parameter $\alpha=0.01$. We note that this $H_{\rm A}$ is large; DFT often overestimates $4f$ magnetocrystalline anisotropy, and recent measurements~\cite{Anisotropy_exp1,Anisotropy_exp2} indicate a strong Ising-like anisotropy with hard-axis saturation fields of at least several tesla (order $5-30\ \mathrm{T})$. These parameters primarily set the electric-field scale (and timescale) required for switching. 
Because the torkance is computed in linear response, the torque field obtained from the fitted VSH coefficients scales linearly with the applied electric field and induces a characteristic precession scale $eEa_0/\hbar$, which competes with the anisotropy precession rate $\gamma\mu_0H_{\rm A}$ in Eq.~\ref{eq:LLG}. We therefore introduce the dimensionless drive parameter $\kappa\equiv eEa_0/(\hbar\gamma \mu_0 H_{\rm A})$ to make this competition explicit. For $\mu_0 H_{\rm A}=47.9\ \mathrm{T}$, the value $\kappa=57.2$ used in Fig.~\ref{fig:TorqueSphereLLG} (c), (d) corresponds to $E\approx6$~V/nm.
Additionally, varying $\alpha$ does not change the qualitative phase portrait: the higher-harmonic torque components still create off-equator fixed points that enable deterministic switching once the threshold condition is met. Additional data for different $\alpha$ are provided in Appendix~\ref{appendix:LLG}.

The resulting spin dynamics are shown in
Fig.~\ref{fig:TorqueSphereLLG}(c,d). For an electric field applied along $\hat{\mathbf{x}}$, a magnetization starting in the northern hemisphere relaxes toward a stable off-equator fixed point below the equator,
while a magnetization starting in the southern hemisphere relaxes toward its mirror partner above the equator. By contrast, when the electric field is increased beyond the deterministic window [Fig.~\ref{fig:TorqueSphereLLG}(b)], the dynamics
enter a strongly precessional regime in which the magnetization oscillates around the fixed points rather than relaxing to them.

To illustrate reversible deterministic control enabled by the symmetry-related off-equatorial fixed points [Fig.~\ref{fig:TorqueSphereLLG}(c), (d)], 
we apply a sequence of two electric-field pulses and
let the magnetization relax in between, as shown in
Fig.~\ref{fig:EfieldOff}. 
In both panels the magnetization is initially
prepared near $+m_z$ and the first pulse $\mathbf{E} \parallel \hat{\mathbf{x}}$ drives it toward an off-equator fixed point in the opposite hemisphere within $10$~ps (fast switching dynamics are shown in insets). 
When the field is turned off, the magnetization relaxes to the $-m_z$ easy axis. A second pulse then
returns the system to the opposite hemisphere, but at different fixed points depending on the pulse polarity. 
In Fig.~\ref{fig:EfieldOff}(a) the two pulses have the same sign, and the magnetization visits two distinct off-equator fixed points that are
symmetrically related as $(m_x,m_y,m_z) \rightarrow (-m_x,m_y,-m_z)$. 
In Fig.~\ref{fig:EfieldOff}(b) the second pulse has the opposite sign (like the conventional symmetry breaking case), and the trajectory instead visits another fixed point related by $(m_x,m_y,m_z) \rightarrow (m_x,-m_y,-m_z)$. 
Thus the dynamics access three symmetry-related fixed points, $(m_x,m_y,m_z)$, $(-m_x,m_y,-m_z)$, and $(m_x,-m_y,-m_z)$, and in all cases the system is deterministically switched between $+m_z$ and $-m_z$.
The key distinction from conventional symmetry-breaking schemes is that both same-polarity and opposite-polarity pulses can realize reversible
switching, while landing in different mirror-related fixed points on the magnetization sphere.

\section{Discussion}\label{sec:discussion}

In this work, we demonstrate a robust and symmetry-guided mechanism for deterministic switching driven by higher-harmonic spin-orbit torques using both toy models and a real material. 
Although PrAlGe is a complex material with strongly
correlated $f$ electrons and multiple electronic bands near the Fermi level, our \emph{ab initio} calculations show that it naturally realizes the key condition identified in the toy model: the higher-order harmonics become comparable in magnitude with the lowest-order torque components that are suppressed over an energy window. In this regime the torque field develops symmetry-allowed off-equator fixed points, enabling deterministic switching of the perpendicular magnetization order parameter without requiring explicit symmetry breaking.

Although we illustrate the switching mechanism using 
$\mu = 0.02$ eV, where the higher-order harmonics are particularly  pronounced, we emphasize that deterministic trajectories are not unique to this choice. As shown in Fig.~\ref{fig:fxnMu}, the relative strength  of higher-order terms varies smoothly with chemical potential, and several nearby values of $\mu$ also yield torque fields with  off-equator fixed points. The mechanism is 
therefore not fine-tuned to a single energy, but persists across a finite window of band filling, further underscoring its robustness.

Because PrAlGe was not selected to maximize higher-harmonic torques, our results should be viewed as a \emph{proof of principle}. The essential switching mechanism is not specific to this material but follows
directly from the angular dependence of the vector spherical harmonics expansion.
The dominance of the $\mathrm{Im}\,\mathbf{Y}^{\mathrm{D}}_{5,5}$ harmonic in
PrAlGe provides one microscopic realization of this mechanism, but the toy-model phase diagrams show that many other combinations of higher-order harmonics can also generate the required fixed-point structure. This suggests that materials with cleaner low-energy electronic structure, reduced multiband complexity, and tunable control over the
balance between conventional and higher-harmonic torques, such that higher-order harmonics remain comparable in magnitude to the lowest-order terms while the overall torque amplitude is large enough to keep the required electric fields moderate, may provide even more favorable platforms. In addition, materials with smaller anisotropy fields would allow the same fixed-point mechanism to operate at reduced electric fields, trading a longer switching timescale for improved experimental accessibility without altering the underlying dynamics.

Experimentally, the mechanism predicts several clear signatures:
(i) nontrivial angular dependence of the spin-orbit torque beyond the standard $\mathbf{p}\times\mathbf{m}$ and $\mathbf{m}\times
(\mathbf{p}\times\mathbf{m})$ forms; (ii) relaxation toward stable opposite-hemisphere fixed points under finite electric field without breaking any in-plane mirror symmetries; and
(iii) reversible switching using electric field pulses of either polarity, with trajectories that access symmetry-related fixed points.
These features distinguish higher-harmonic-driven switching from conventional symmetry-breaking schemes and may be accessible in optical pump-probe measurements, ultrafast Hall-based torque probes, or
spin-torque ferromagnetic resonance experiments.

We also note that strong spin-orbit torques can modify the electronic topology itself. When the applied electric field rotates the magnetization away from its equilibrium orientation, the rotation symmetries that make the pairs of Weyl nodes energy degenerate may no longer be preserved (See Appendix~\ref{appendix:WP}). Thus, a finite electric field can shift the Weyl-node positions in both momentum and energy, dynamically tuning the Berry curvature landscape during the switching process. Although this effect is not the origin of the deterministic switching identified here, it provides an additional experimentally accessible signature of higher-harmonic torques and may offer another route to materials optimization: materials in which the Weyl nodes are close to the Fermi level or highly sensitive to electric-field-induced spin tilting may exhibit an even stronger higher-harmonic torque response.

Future work may focus on identifying materials in which higher-order harmonics are intrinsically enhanced—such as noncentrosymmetric
semimetals with simpler band topology, magnetic altermagnets with strong anisotropic Berry curvature responses~\cite{Altermagnet_Review_PRX}, or engineered heterostructures where symmetry and band filling can be tuned. 
The toy-model framework introduced here provides a systematic strategy for such a search by connecting crystalline symmetry, torque harmonics, and switching
phase diagrams. As a whole, our findings establish higher-harmonic spin-orbit torques as an underexplored but powerful route for electric-field control of magnetization.

\section{Acknowledgment}
The authors would like to acknowledge Dr. Jia Shi and Dr. Rajibul Islam for their assistance and helpful discussions throughout this project. 
The work done at University of Alabama at Birmingham is supported by the National Science Foundation under Grant No. OIA-2229498, UAB internal startup funds, and UAB Faculty Development Grant Program, Office of the Provost. The authors gratefully acknowledge the resources provided by the University of Alabama at Birmingham IT-Research Computing group for high performance computing (HPC) support and CPU time on the Cheaha compute cluster.
This work used Stampede3 at Texas Advanced Computing Center through allocation PHY250049 from the Advanced Cyberinfrastructure Coordination Ecosystem: Services \& Support(ACCESS) program, which is supported by U.S. National Science Foundation grants \#2138259, \#2138286, \#2138307, \#2137603, and \#2138296.\ \\

\appendix
\section{Symmetry Analysis}\label{appendix:SymmAnl}
This appendix supplements Sec.~\ref{sec:symm} by outlining the vector spherical
harmonics (VSH) expansion used to derive the symmetry-allowed forms of spin-orbit
torque (SOT) in systems with $C_{4z}$ rotational symmetry. The methodology closely
follows Ref.~\cite{Xue2023}. 
For a magnetization direction $\mathbf{\hat{m}}=(\sin\theta \cos\phi, \sin\theta \sin\phi, \cos\theta)$, the vector spherical harmonics components are defined in terms of scalar spherical harmonics $Y_{lm}(\mathbf{\hat{m}})$ as
	\begin{align}
		& \mathbf{Y}^{\rm D}_{lm}(\mathbf{\hat{m}})
		=\frac{\nabla_{\mathbf{\hat{m}}} Y_{lm}(\mathbf{\hat{m}})}
		{\sqrt{l(l+1)}},\label{eq:DampingVSH}\\
		&\mathbf{Y}^{\rm F}_{lm}(\mathbf{\hat{m}})
		=\frac{\mathbf{\hat{m}}\times\nabla_{\mathbf{\hat{m}}} Y_{lm}(\mathbf{\hat{m}})}
		{\sqrt{l(l+1)}}.\label{eq:FieldVSH}
	\end{align}
Both  $\mathbf{Y}^{\rm D}_{lm}$ and $\mathbf{Y}^{\rm F}_{lm}$ are orthogonal to $\hat{\mathbf{m}}$  and are classified as dampinglike and fieldlike respectively based on their roles in spin dynamics governed by LLG equation. The effective fields associated with the fieldlike $\mathbf{Y}^{\rm F}$ terms are curl-free and correspond to gradients of scalar fields,
while the dampinglike terms are generated from their curls and can be written as
$\mathbf{\hat{m}}\times\mathbf{Y}^{\rm F}_{lm}$.

Time-reversal symmetry (TRS) imposes constraints on which harmonics can appear.
As summarized in Table~\ref{table:vsh},
$\mathbf{Y}^{\rm D}_{lm}$ is odd (even) under TRS for even (odd) $l$,
while $\mathbf{Y}^{\rm F}_{lm}$ is even (odd) under TRS for even (odd) $l$.

\begin{table}[htbp]
		\centering
		\begin{tabular}{| c | c| c |}
			\hline
			& $l$ even & $l$ odd \\ \hline
			$\mathbf{Y}^{\rm D}_{lm}$ & odd & even \\ \hline
			$\mathbf{Y}^{\rm F}_{lm}$ & even & odd \\  \hline  
		\end{tabular}
            \caption{Time-reversal symmetry properties of the vector spherical harmonics $\mathbf{Y}^{\rm D,F}_{lm}$.}
            \label{table:vsh}
	\end{table}

The SOT can be expressed in linear response as
\begin{equation}
\boldsymbol{\mathcal{T}}_{\mathbf{\hat{E}}}(\mathbf{\hat{m}})
=\boldsymbol{\tau}_{\mathbf{\hat{E}}}(\mathbf{\hat{m}})\mathbf{E} ,
\label{eq:tensor}
\end{equation}
where the torkance tensor is expanded in the VSH basis:
\begin{equation}
\boldsymbol{\tau}_{\mathbf{\hat{E}}}(\mathbf{\hat{m}})
=\sum_{lm}\Big[\mathbf{Y}^{\rm D}_{lm}{C}^{\rm D}_{lm}({\mathbf{\hat{E}}})
+\mathbf{Y}^{\rm F}_{lm}{C}^{\rm F}_{lm}({\mathbf{\hat{E}}})\Big].
\end{equation}
Here the coefficients $C^{\rm D,F}_{lm}$ represent the contributions of the
corresponding VSH terms to the torque.

For a system with $C_{4z}$ rotational symmetry, the relevant symmetries include
the fourfold rotation about $\hat{\mathbf{z}}$ together with mirror planes in both the
$xz$ and $yz$ directions. The fourfold rotation ensures that the torkance is invariant when we perform a rotation angle $\gamma=\pi/2$ from $x$($y$)-axis~\cite{Xue2023}:

\begin{equation}
		\label{eq:rotation_phase}
  		\begin{split}
		&C_{lm}(\mathbf{\hat{\mathbf{x}}})e^{-\text{i}m\gamma}=C_{lm}(\mathbf{\hat{\mathbf{x}}})\cos{\gamma}+C_{lm}(\mathbf{\hat{\mathbf{y}}})\sin{\gamma},\\
            &C_{lm}(\mathbf{\hat{\mathbf{y}}})e^{-\text{i}m\gamma}=-C_{lm}(\mathbf{\hat{\mathbf{x}}})\sin{\gamma}+C_{lm}(\mathbf{\hat{\mathbf{y}}})\cos{\gamma}.
	    \end{split}  
        \end{equation}
These equations require that only harmonics with
$m=4n\pm 1$ are symmetry-allowed. Terms with $m=4n$ or $m=4n+2$ are forbidden,
since they acquire inconsistent phase factors under a 90° rotation and thus do
not remain invariant. 
Consequently, the torque expansions for $\mathbf{E}\!\parallel\!\hat{\mathbf{x}}$ and
$\mathbf{E}\!\parallel\!\hat{\mathbf{y}}$ involve the same set of harmonics
($m=4n\pm 1$), differing only by the phase relation enforced by the
$C_{4z}$ rotation:
\begin{equation}
		\label{eq:EyandEx}
		C_{l,4n\pm1}(\mathbf{\hat{\mathbf{y}}})=\mp i C_{l,4n\pm1}(\mathbf{\hat{\mathbf{x}}}).
    \end{equation}

The mirror planes impose specific constraints on the allowed vector spherical
harmonics. These mirror-plane constraints are summarized in Table~\ref{table:symmetryEx},
which lists the allowed $(l,m)$ values for the real and imaginary parts of
$\mathbf{Y}^{\rm D,F}_{lm}$. The coordinate system is chosen such that
$\hat{\mathbf{x}}\parallel\hat{\mathbf{E}}$, $\hat{\mathbf{z}}\parallel\hat{\mathbf{n}}$ (film normal), and
$\hat{\mathbf{y}}=\hat{\mathbf{n}}\times\hat{\mathbf{E}}$.
In practice, we work with real-valued vector spherical harmonics obtained from
the real and imaginary parts of $\mathbf{Y}_{lm}$. This guarantees that all basis
functions are real and orthonormal on the unit sphere. In this convention $m$
runs from $0$ to $l$, with $\text{Re}\,\mathbf{Y}_{lm}$ and $\text{Im}\,\mathbf{Y}_{lm}$ providing
two independent real basis functions for each nonzero $m$. This is the reason
why the mirror-plane constraints in Table~\ref{table:symmetryEx} are expressed
separately for the real and imaginary parts of $\mathbf{Y}^{\rm D,F}_{lm}$.

\begin{table}[htbp]
\centering
\begin{tabular}{| c | c| c |}
\hline
Mirror plane &  \text{Re}$\mathbf{Y}^{\rm D,F}_{lm}$ & \text{Im}$\mathbf{Y}^{\rm D,F}_{lm}$ \\ \hline
$\sigma_{{\hat{\mathbf{E}}},{\hat{\mathbf{n}}}}$ & $l$ even & $l$ odd \\  \hline 			
$\sigma_{{\hat{\mathbf{p}}},{\hat{\mathbf{n}}}}$ & $l+m$ odd & $l+m$ even \\ \hline
$\sigma_{{\hat{\mathbf{p}}},{\hat{\mathbf{E}}}}$ & $m$ even  & $m$ even \\ \hline
\end{tabular}
\caption{Symmetry constraints on $(l,m)$ imposed by different mirror planes.
If multiple mirrors are present, their constraints combine.}
\label{table:symmetryEx}
\end{table}

By combining the mirror constraints in Table~\ref{table:symmetryEx} with the
time-reversal properties in Table~\ref{table:vsh}, we obtain the symmetry-allowed
forms of the spin-orbit torque for $E\!\parallel\!\hat{\mathbf{x}}$, given in
Eqs.~\ref{eq:tau_even} and \ref{eq:tau_odd} of the main text.

\begin{equation}
\begin{aligned}
\boldsymbol{\tau}^{\text{even}}_{\hat{\mathbf{x}}} = \sum_{l,m}
[ &C^{\rm F}_{2l,4m\pm1}\operatorname{Re}\mathbf{Y}^{\rm F}_{2l,4m\pm1} \\
+&C^{\rm D}_{2l+1,4m\pm1}\operatorname{Im}\mathbf{Y}^{\rm D}_{2l+1,4m\pm1}],
\end{aligned}
\label{eq:tau_eveny}
\end{equation}

\begin{equation}
\begin{aligned}
\boldsymbol{\tau}^{\text{odd}}_{\hat{\mathbf{x}}} = \sum_{l,m} [ &C^{\rm D}_{2l,4m\pm1}\operatorname{Re}\mathbf{Y}^{\rm D}_{2l,4m\pm1} \\
+&C^{\rm F}_{2l+1,4m\pm1}\operatorname{Im}\mathbf{Y}^{\rm F}_{2l+1,4m\pm1}].
\end{aligned}
\label{eq:tau_oddy}
\end{equation}

By utilizing Eq.~\ref{eq:EyandEx}, we obtain
\begin{equation}
\begin{aligned}
\boldsymbol{\tau}^{\text{even}}_{\hat{\mathbf{y}}} = \sum_{l,m}
[ &\pm C^{\rm F}_{2l,4m\pm1}\rm{Im}\mathbf{Y}^{\rm F}_{2l,4m\pm1} \\
&\mp C^{\rm D}_{2l+1,4m\pm1}\rm{Re}\mathbf{Y}^{\rm D}_{2l+1,4m\pm1}],
\end{aligned}
\label{eq:tau_eveny}
\end{equation}

\begin{equation}
\begin{aligned}
\boldsymbol{\tau}^{\text{odd}} _{\hat{\mathbf{y}}} = \sum_{l,m} [ &\pm C^{\rm D}_{2l,4m\pm1}\rm{Im}\mathbf{Y}^{\rm D}_{2l,4m\pm1} \\
&\mp C^{\rm F}_{2l+1,4m\pm1}\rm{Re}\mathbf{Y}^{\rm F}_{2l+1,4m\pm1}].
\end{aligned}
\label{eq:tau_oddy}
\end{equation}

\section{Critical Field LLG Simulation}\label{appendix:LLG}

To demonstrate that the higher-harmonic torque consistently drives magnetization
to the same off-equator fixed point, and to confirm that this outcome is not
merely coincidental, we conducted additional Landau-Lifshitz-Gilbert (LLG)
simulations starting from multiple randomly chosen initial conditions near the
easy-axis poles. As shown in Table~\ref{tab:simulation_data_big}, all tested
initial magnetization directions consistently evolve to the same stable fixed
point when higher-harmonic torque contributions are sufficiently large
(ratio = 0.4) shown in table ~\ref{tab:simulation_data_big}. In contrast, with smaller higher-harmonic torque contributions
(ratio = 0.1) shown in table ~\ref{tab:simulation_data_small}, the magnetization is driven only to the equator,
reflecting nondeterministic outcomes. These results confirm that the
deterministic switching enabled by higher-harmonic torques is robust and occurs
systematically, provided the initial magnetization is away from the equatorial
symmetry point.

\begin{table}[htbp!]
\centering
\begin{tabular}{ccc}
\hline
$\theta$ & $\phi$ & final $m_z$ \\
\hline
0.361 & 4.476 & -0.086 \\
0.380 & 0.993 & -0.086 \\
0.337 & 3.793 & -0.086 \\
0.216 & 5.659 & -0.086 \\
0.260 & 2.176 & -0.086 \\
0.028 & 4.043 & -0.086 \\
0.072 & 0.560 & -0.086 \\
0.116 & 3.360 & -0.086 \\
0.117 & 5.226 & -0.086 \\
0.073 & 1.743 & -0.086 \\
0.029 & 4.543 & -0.086 \\
0.261 & 0.127 & -0.086 \\
\hline
\end{tabular}
\caption{LLG simulation results for  $E=0.8 {H_{\rm A}}$ and $\rm{ratio} = 0.4$.
Initial conditions ($0<\theta<\pi/8$, $0<\phi<2\pi$) all converge to the
same off-equator fixed point.}
\label{tab:simulation_data_big}
\end{table}

\begin{table}[htbp!]
\centering
\begin{tabular}{ccc}
\hline
$\theta$ & $\phi$ & final $m_z$ \\
\hline
0.361 & 0.542 & 0.000 \\
0.317 & 3.342 & 0.000 \\
0.236 & 5.208 & 0.000 \\
0.280 & 1.725 & 0.000 \\
0.048 & 3.592 & 0.000 \\
0.091 & 0.108 & 0.000 \\
0.135 & 2.908 & 0.000 \\
0.097 & 4.775 & 0.000 \\
0.053 & 1.292 & 0.000 \\
0.286 & 3.158 & 0.000 \\
0.242 & 5.958 & 0.000 \\
0.311 & 1.542 & 0.000 \\
\hline
\end{tabular}
\caption{LLG simulation results for  $E=0.8 {H_{\rm A}}$ and $\rm{ratio} = 0.1$.
All initial conditions converge to the equator, leading to nondeterministic
outcomes.}
\label{tab:simulation_data_small}
\end{table}

\begin{figure}[htbp]
    \centering
    \includegraphics[width=0.4\textwidth]{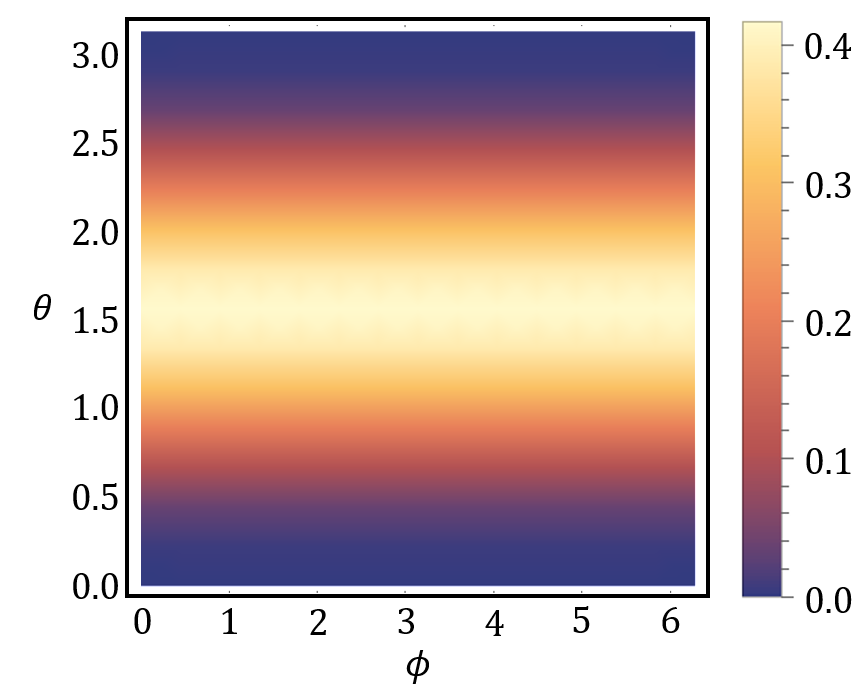} 
    \caption{Energy-like scalar potential $\Phi(\theta,\phi)\propto \mathrm{Im}\,Y_{3,3}(\theta,\phi)$ associated with the $\ell=3,m=3$ fieldlike harmonic. The pattern is symmetric under $\theta\rightarrow \pi-\theta$ (equivalently $m_z\rightarrow -m_z$), indicating no north-south energetic bias.}    \label{fig:energyLandscape} 
\end{figure}

We emphasize that the switching enabled by higher-harmonic torques is not due to a north–south energetic bias because mirror symmetry is not broken. For a field-like harmonic, one can associate an energy-like scalar potential $\Phi(\mathbf{\hat{m}})$ on the unit sphere such that the corresponding fieldlike torque has the form 
$T^{\rm F} \propto \mathbf{\hat{m}} \times \nabla_{\hat{\mathbf{m}}} \Phi(\mathbf{\hat{m}})$. For $\text{Im}\mathbf{Y}^{\rm F}_{3,3}$ in Eq.~\ref{eq:toymodel}, $\Phi$ is proportional (up to a constant factor) to the scalar spherical harmonics $\rm{Im}Y_{3,3}(\theta,\phi)$. Importantly, $\rm{Im}Y_{3,3}(\theta,\phi)$ is symmetric under $\theta\rightarrow \pi-\theta$ (or equivalently $m_z\rightarrow-m_z$), and therefore cannot generate $m_z$-odd north-south energy bias. 
This is also evident from the explicit Cartesian form: $\Phi\propto
\frac{ m_y \left( 3 m_x^2 - m_y^2 \right) }{ |\mathbf{m}|^3 }$, which is an even function of $m_{z}$. 
Figure~\ref{fig:energyLandscape} visualizes $\Phi(\theta,\phi)$ and makes the north-south symmetry explicit.

To complement the toy-model robustness tests above, we performe the same type of initial-condition sampling using the \textit{ab initio} torque field [Fig.~\ref{fig:TorqueSphereLLG}(a)]. Specifically, we initialize the magnetization at multiple randomly chosen directions within a small cone around the $+\hat{z}$ easy-axis pole, $0<\theta<\pi/100$ and $0<\phi<2\pi$, and integrate the LLG dynamics for each case. In all tested samples, the trajectories relax to the same stable off-equatorial attractor with $m_z\simeq -0.147$ (Table~\ref{tab:final_mz}). This confirms that the deterministic switching obtained from the first-principles torque field persists over a finite neighborhood of initial conditions and does not require a fine-tuned starting direction.

\begin{table}[htbp]
\centering
\begin{tabular}{ccc}
\hline
$\theta$ & $\phi$ & Final $m_z$ \\
\hline
0.014 & 2.268 & -0.147 \\
0.025 & 4.687 & -0.147 \\
0.008 & 1.574 & -0.147 \\
0.011 & 1.489 & -0.147 \\
0.013 & 3.537 & -0.147 \\
0.029 & 4.962 & -0.147 \\
0.018 & 2.254 & -0.147 \\
0.011 & 2.495 & -0.147 \\
0.031 & 0.088 & -0.147 \\
0.028 & 3.835 & -0.147 \\
0.021 & 4.302 & -0.147 \\
\hline
\end{tabular}
\caption{LLG simulation for initial conditions $0<\theta<\frac{\pi}{100}, 0<\phi<2\pi$ for ab initio torque results}
\label{tab:final_mz}
\end{table}

The relevant control parameter is the dimensionless drive strength relative to the perpendicular anisotropy field $\kappa\equiv eEa_0/(\hbar\gamma \mu_0 H_{\rm A})$. For a fixed \textit{ab initio } torque field and fixed $H_{\rm A}$, the critical drive also depends on Gilbert damping $\alpha$. In particular, increasing $\alpha$ increases the required threshold, as summarized in Table~\ref{tab:alpha_damping}.

\begin{table}[htbp]
\centering
\begin{tabular}{cccc}
\hline
$\alpha$ & Critical $E$ (V/nm) & $\kappa$ & Final $m_z$ \\
\hline
0.01 & 5.8 & 55.30 & -0.145 \\
0.02 & 6.0 & 57.20 & -0.145 \\
0.03 & 6.2 & 59.11 & -0.145 \\
0.04 & 6.4 & 61.01 & -0.145 \\
\hline
\end{tabular}
\caption{Dependence of critical switching field, $\kappa$, and final magnetization $m_z$ on Gilbert damping $\alpha$.}
\label{tab:alpha_damping}
\end{table}

\section{Magnetization Dynamics Phase Diagrams for $\text{Im}  \mathbf{Y}^{\rm D}_{3,3}$ and $\text{Im}  \mathbf{Y}^{\rm D}_{5,5}$ }\label{appendix:LLGPhaseDiagrams}

Section~\ref{sec:toymodel} studies the magnetization dynamics of a toy model incorporating conventional terms with the addition of the higher-harmonic term $\text{Im}  \mathbf{Y}^{\rm F}_{3,3}$. Utilizing Landau-Lifshitz dynamics, Fig.~\ref{fig:2} (c) shows that this higher-harmonic term enables deterministic switching when its contribution is large enough. Here we apply the same LLG-based analysis to two
additional higher-order harmonics that arise in the \emph{ab initio}
torques of PrAlGe, namely
$\mathrm{Im}\,\mathbf{Y}^{\rm D}_{3,3}$ and
$\mathrm{Im}\,\mathbf{Y}^{\rm D}_{5,5}$.

\begin{figure}[htbp]
    \centering
    \includegraphics[width=0.5\textwidth]{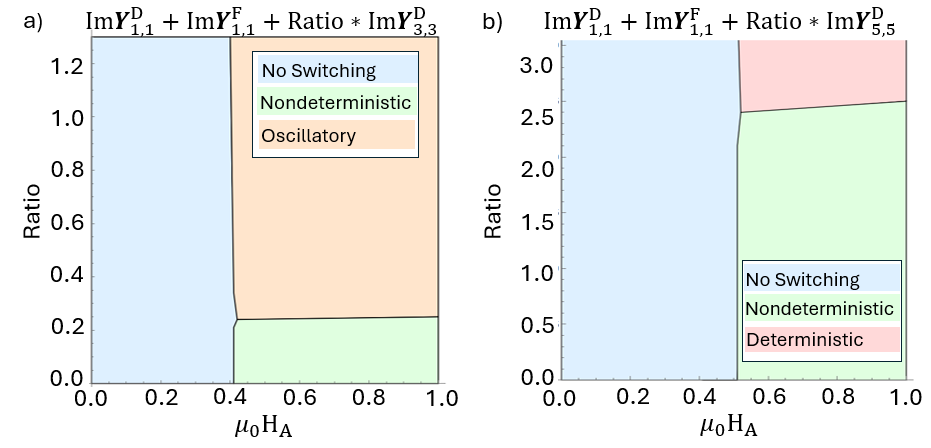} 
    \caption{Phase diagrams of magnetization dynamics for higher-order torque harmonics: (a) $\mathrm{Im}\,\mathbf{Y}^{\rm D}_{3,3}$
    and (b) $\mathrm{Im}\,\mathbf{Y}^{\rm D}_{5,5}$.}
    \label{fig:HigherOrderPhase} 
\end{figure}

As shown in Fig.~\ref{fig:HigherOrderPhase}(a), the term $\mathrm{Im}\,\mathbf{Y}^{\rm D}_{3,3}$ does not produce deterministic switching but instead generates an extended oscillatory
(precessional) regime. Such oscillatory dynamics can also be useful as nano-oscillators, in which the frequency can be tuned by varying the ratio of higher-harmonic to conventional torque amplitudes~\cite{Xue2021SOT_AFM}.

Figure~\ref{fig:HigherOrderPhase}(b) presents the phase diagram for $\mathrm{Im}\,\mathbf{Y}^{\rm D}_{5,5}$, which is identified in Sec.~\ref{sec:firstprinc} as the dominant higher-harmonic term responsible for deterministic switching in PrAlGe. In this case, a sufficiently large ratio of $\mathrm{Im}\,\mathbf{Y}^{\rm D}_{5,5}$ to the
lowest-order torque components produces the same off-equator fixed-point structure as the toy model Eq.~\ref{eq:toymodel}, enabling deterministic switching. The
required ratio is larger than for the $\mathrm{Im}\,\mathbf{Y}^{\rm F}_{3,3}$ 
term considered in the main text, but the qualitative mechanism is similar. 

\section{First-Principles Details of Torque Evaluations}\label{appendix:FPC}
PrAlGe is modeled using Vienna Ab initio Simulation Package utilizing a GGA$+U$ method. The lattice constants for PrAlGe used in this calculation are a=4.24622\AA{}, c=14.6421\AA{}. We choose the primitive lattice vectors as
\begin{align*}
\mathbf{a}_1 &= (-a,\; a,\; c)/2, \\
\mathbf{a}_2 &= ( a,\; -a,\; c)/2, \\
\mathbf{a}_3 &= ( a,\; a,\; -c)/2.
\end{align*}
We utilize a $14 \times 14 \times 14$ Gamma-centered mesh and an energy cutoff of $500$~eV. We implement a Hubbard $U$ of 4~eV, consistent with prior work ~\cite{Chang2018}~\cite{Sanchez2020}. Both collinear and noncollinear calculations are performed.The k-path used to obtain the band structures in Figs. ~\ref{fig:3}(c) and (d) follows the high symmetry points reported in Table ~\ref{table:highsymmpoints}.

\begin{table}[h!]
\centering
\begin{tabular}{lccc}
\hline
Label & $k_1$ & $k_2$ & $k_3$ \\
\hline
$\Gamma$ & 0.0000 & 0.0000 & 0.0000 \\
$\Sigma_0$    & -0.2710 & 0.2710 & 0.2710 \\
N        & 0.0000 & 0.5000 & 0.0000 \\
$\Sigma$        & 0.2710 & 0.7290 & -0.2710 \\
Z        & 0.5000 & 0.5000 & -0.5000 \\
X        & 0.0000 & 0.0000 & 0.5000 \\
\hline
\end{tabular}
\caption{High-symmetry k-points used as the path for the band structure of PrAlGe in fractional coordinates.}
\label{table:highsymmpoints}
\end{table}

From the plane-wave calculation, we obtain a tight-binding Hamiltonian of the collinear calculation using Wannier90~\cite{Wannier90}. The atomic orbital basis used are Ge: $s,p$; Al: $s,p$; Pr: $d,f$, and these are chosen from the density of states calculations and agree with atomic orbitals chosen in previous literature of this material~\cite{Chang2018}~\cite{Sanchez2020}. Now that a tight-binding Hamiltonian has been achieved, WannSymm~\cite{Zhi2022WannSymm} is then used to symmetrize the Hamiltonian. 

The \textit{ab initio} torkance entering Eqs.\ref{eq:eventorkance} and \ref{oddtorkance} is evaluated on a 100 x 100 x 100 k point mesh. This choice is guided by explicit convergence tests performed at a representative magnetization direction $(\theta, \phi)=(0,0)$. The selected mesh provides a practical compromise between numerical convergence and computational cost.

\begin{figure}[htbp]
    \centering
    \includegraphics[width=0.5\textwidth]{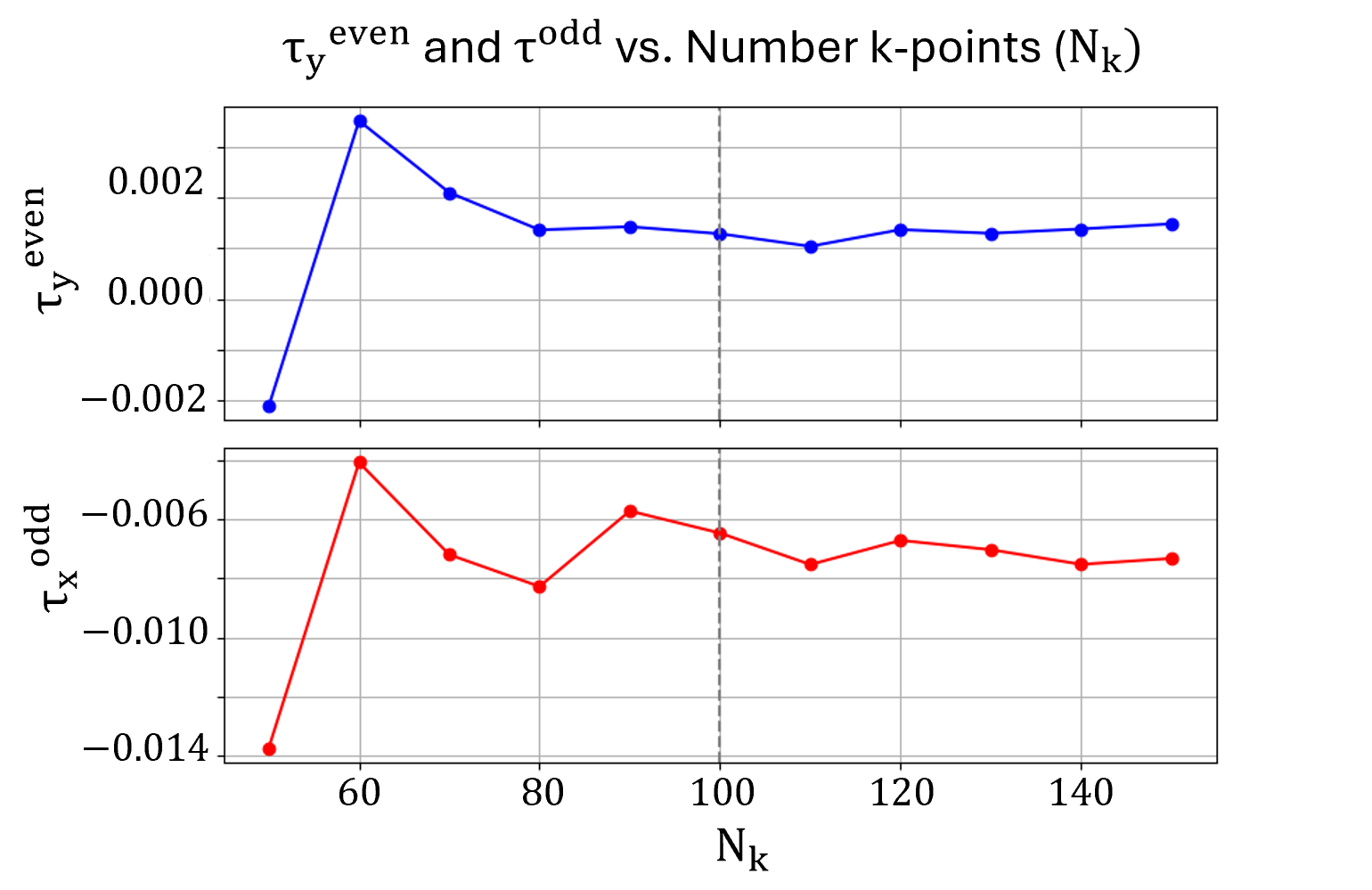} 
\caption{Convergence of representative even and odd torkance components as a function of the total number of k-points $N_k$ (evaluated at a fixed magnetization direction, $\theta=0$ and $\phi=0$). The vertical dashed line marks the k-mesh used for the production torkance calculations.}
\label{fig:kmeshConv} 
\end{figure}

\section{Fitting at the Fermi Energy}\label{appendix:FermiFit}

Spin-orbit torques were computed over a range of chemical potentials, as shown in Fig.~\ref{fig:fxnMu}. For completeness, we also provide the fitted vector spherical harmonics coefficients evaluated at the DFT Fermi level $\mu=0$, summarized in Table.~\ref{table:twopart}.

\begin{table}[h!]
\centering

\begin{tabular}{| c | c | c | c | c | c | c | c |}
\hline
$C^{\rm D}_{1,1}$ &
$C^{\rm D}_{3,3}$ &
$C^{\rm D}_{5,1}$ &
$C^{\rm D}_{5,3}$ &
$C^{\rm D}_{5,5}$ &
$C^{\rm D}_{7,1}$ &
$C^{\rm F}_{2,1}$ &
$C^{\rm F}_{4,3}$ \\ \hline
-0.008 & -0.006 & -0.006 & 0.003 & 0.006 & 0.001 & 0.003 & 0.005 \\ \hline
\end{tabular}

\vspace{0.6cm}

\begin{tabular}{| c | c | c | c | c | c |}
\hline
$C^{\rm D}_{2,1}$ &
$C^{\rm D}_{4,1}$ &
$C^{\rm D}_{4,3}$ &
$C^{\rm F}_{1,1}$ &
$C^{\rm F}_{3,1}$ &
$C^{\rm F}_{3,3}$ 
\\ \hline
0.023 & -0.001 & -0.001 & -0.026 & -0.001 & 0.001   \\ \hline
\end{tabular}

\caption{Fitting coefficients for spin-orbit torque at the Fermi energy $\mu=0$ eV with units $e a_0/\hbar$. 
The upper table includes the time-reversal even terms and the lower table are the time-reversal odd. Terms smaller than 0.001 are excluded for brevity.}
\label{table:twopart}
\end{table}

\section{Weyl Points}\label{appendix:WP}
WannierTools ~\cite{WU2017} was used to identify Weyl points for the noncollinear tight-binding model derived from  DFT, as well as the collinear tight-binding model with atomically added spin-orbit coupling (SOC).

Table~\ref{table:WPs_noncol} lists the seven independent Weyl points found in the noncollinear model. Accounting for crystal symmetries, these correspond to a total of 56 Weyl nodes in the Brillouin zone, consistent with earlier reports~\cite{Chang2018}.

\begin{table}
\begin{tabular}{cccc}
\toprule
$k_x$ (1/\AA) & $k_y$ (1/\AA) & $k_z$ (1/\AA)& Energy (meV) \\
\midrule
0.366	&0.058	&0.280  & -8.1
 \\
0.287	&0.245  &0.054	& 28.9
 \\
0.236	&0.291  &0.048	& 31.1
 \\
0.013	&0.688	&0.391	& 46.5
 \\
0.039	&0.381  &0.289	& 50.4
 \\
0.230	&0.309  &0.001	& 50.9
 \\
0.416	&0.254  &0.004  & 69.6
 \\
\bottomrule
\end{tabular}
\caption{Symmetry-inequivalent Weyl nodes for the noncollinear tight-binding Hamiltonian with magnetization along $\hat{\mathbf{z}}$. Energies are measured relative to the Fermi level.}

\label{table:WPs_noncol}
\end{table}
After including atomic SOC in the collinear tight-binding Hamiltonian, six independent Weyl nodes are obtained, listed in Table~\ref{table:WPs_atom}. The reduction from seven to six inequivalent nodes arises because two nearby Weyl points in the noncollinear case merge into one under the collinear+SOC treatment.

\begin{table}
\begin{tabular}{cccc}
\toprule
$k_x$ (1/\AA) & $k_y$ (1/\AA) & $k_z$ (1/\AA)& Energy (meV) \\
\midrule
-0.002	&0.644 &0.412 &-21.4
 \\
0.289	&0.245 &0.045 &2.6
 \\
0.036	&0.338 &0.349 &30.8
 \\
0.386	&0.051 &0.299 &36.3
 \\
0.257	&0.431 &0.003 &48.3
 \\
0.019	&0.315 &0.396 &83.0
 \\
\bottomrule
\end{tabular}
\caption{Symmetry-inequivalent Weyl nodes for the collinear tight-binding Hamiltonian with atomic SOC. Energies are measured relative to the Fermi level.}
\label{table:WPs_atom}
\end{table}

A direct comparison between the two models is presented in Table~\ref{tab:NC_vs_SOC}. Most Weyl nodes in the noncollinear model map cleanly to those in the collinear+SOC model, with small momentum displacements $\Delta k \lesssim 0.05$-$0.13~1/\text{\AA}$. One noncollinear Weyl point ($k \approx 0.236,0.291,0.048$) merges with its neighbor after SOC is added, explaining the reduction in the number of inequivalent nodes. Overall, the two approaches yield a consistent Weyl topology, with only minor differences in precise energies and positions.

\begin{table}[htbp]
\centering
\begin{tabular}{ccc|ccc|c}
\hline
\multicolumn{3}{c|}{Noncollinear}  &
\multicolumn{3}{c|}{Collinear+SOC} & $\Delta k$ ($1/\text{\AA}$) \\
$k_x$ & $k_y$ & $k_z$ &   $k_x$ & $k_y$ & $k_z$ &  \\
\hline
0.366 & 0.058 & 0.280 &  0.386 & 0.051 & 0.299 & 0.030 \\
0.287 & 0.245 & 0.054 &   0.289 & 0.245 & 0.045 & 0.010 \\
0.236 & 0.291 & 0.048 &   0.245 & 0.289 & -0.045 & 0.010 \\
0.013 & 0.688 & 0.391 & -0.002 & 0.644 & 0.412 & 0.046 \\
0.039 & 0.381 & 0.289 &   0.036 & 0.338 & 0.349 & 0.074 \\
0.230 & 0.309 & 0.001 &   0.257 & 0.431 & 0.003 & 0.127 \\
0.416 & 0.254 & 0.004 &   0.431 & 0.257 & -0.003 & 0.017 \\
\hline
\end{tabular}
\caption{Comparison of symmetry-inequivalent Weyl nodes in the noncollinear vs collinear+SOC models. Coordinates are given in reciprocal Cartesian units ($1/\text{\AA}$). The last column reports the momentum displacement $\Delta k$ between corresponding nodes.}
\label{tab:NC_vs_SOC}
\end{table}

From these six independent Weyl points, a total of 48 symmetry-related Weyl points emerge due to the system's crystal symmetries. Specifically, the system's fourfold rotational symmetry around the $z$-axis ($C_{4z}$) generates three additional Weyl points from each independent one through 90° rotations in momentum space. Consequently, groups of four Weyl points share the same chirality.

To satisfy the Nielsen-Ninomiya theorem~\cite{NIELSEN1983,Armitage2018RMP}, which demands zero net chirality in periodic crystal, each Weyl point must possess a partner with opposite chirality. This partner arises from the combined action of mirror symmetry about the $yz$-plane ($\sigma_{yz}$) and time-reversal symmetry. Mirror symmetry preserves momentum while reversing spin, thus reversing chirality. Conversely, time-reversal symmetry simultaneously reverses both spin and momentum, leaving chirality unchanged. Thus, the combination of mirror and time-reversal symmetry ensures each Weyl point has a partner of opposite chirality.

As a representative example, Table~\ref{table:WeylPoints} lists the full set of eight Weyl points (four pairs with opposite chirality) generated from the Weyl point located near the Fermi level at $\mathbf{k} = (0.245, 0.289, -0.045)$ with an energy of $2.6$~meV.

\begin{table}[htbp]
\centering
\begin{tabular}{ccccc}
\hline
$k_x$ (1/\AA) & $k_y$ (1/\AA) & $k_z$ (1/\AA) & Chirality & E (meV) \\
\hline
 0.245  & 0.289  & -0.045  &  1 & 2.6\\
-0.289  &  0.245  & -0.045  &  1 & 2.6\\
-0.245  & -0.289  & -0.045  &  1 & 2.6\\
0.289  & -0.245  & -0.045  &  1 & 2.6\\
 0.245  & -0.289  &  0.045  & -1 & 2.6\\
-0.289  & -0.245  &  0.045  & -1 & 2.6\\
-0.245  & 0.289  &  0.045  & -1 & 2.6\\
0.289  &  0.245  &  0.045  & -1 & 2.6\\
\hline
\end{tabular}
\caption{Representative set of symmetry-generated Weyl nodes. The first four are related by $C_{4z}$ and share identical chirality. The last four are obtained via $\sigma_{yz}$ and time-reversal, yielding opposite chirality.}
\label{table:WeylPoints}

\end{table}

Once the magnetization is rotated to $\hat{\mathbf{m}}\parallel\hat{\mathbf{y}}$, $C_{4z}$ symmetry of the $\hat{\mathbf{z}}$-aligned state is lifted, and the Weyl nodes shift in both momentum and energy. Specifically, the set of eight energy-degenerate nodes at $\hat{\mathbf{m}}\parallel\hat{\mathbf{z}}$ splits into four energy-nondegenerate nodes; each of these generates a quartet of symmetry-related partners under the operations that remain at $\hat{\mathbf{m}}\parallel\hat{\mathbf{y}}$ (notably $\sigma_{xz}$, $\sigma_{yz}\mathcal{T}$, and $C_{2z}\mathcal{T}$). A representative set of the four independent nodes is listed in Table~\ref{table:WeylPoints_My}. Relative to the $\hat{\mathbf{m}}\parallel\hat{\mathbf{z}}$ case, the energy shifts exceed $\sim 35$ meV and the momentum displacements are $\sim 0.02~\text{\AA}^{-1}$(typical).

\begin{table}[htbp]
\centering
\begin{tabular}{ccccc}
\hline
$k_x$ (1/\AA) & $k_y$ (1/\AA) & $k_z$ (1/\AA) & Chirality & E (meV) \\
\hline
 0.257  &  0.283  &  -0.035  & 
 1 & 33.4\\
 -0.278  & 0.258  &  -0.037  &  1 & -19.7\\
-0.241  &  -0.304  & -0.059  &  1 & -10.3\\
 0.309  &  -0.238  & -0.050  &
 1 & 38.0\\
\hline
\end{tabular}
\caption{Four \emph{independent} Weyl nodes for $\hat{\mathbf{m}}\parallel\hat{\mathbf{y}}$ corresponding to the first four $C_{4z}$ symmetry-equivalent nodes in Table.~\ref{table:WeylPoints}. Each entry generates a fourfold set of symmetry partners related by the operations that remain at $\hat{\mathbf{m}}\parallel\hat{\mathbf{y}}~(\sigma_{xz}, \sigma_{yz}\mathcal{T}$, and $C_{2z}\mathcal{T}$). The corresponding partners share the same energy; chirality transforms according to the underlying symmetry (in particular, it is preserved under $C_{2z}\mathcal{T}$ and changes sign under the mirrors).}
\label{table:WeylPoints_My}
\end{table}

\section{Tuning Spin-orbit Coupling Strength}\label{appendix:tuneSOC}

\begin{figure*}[htbp]
    \centering
    \includegraphics[width=\textwidth]{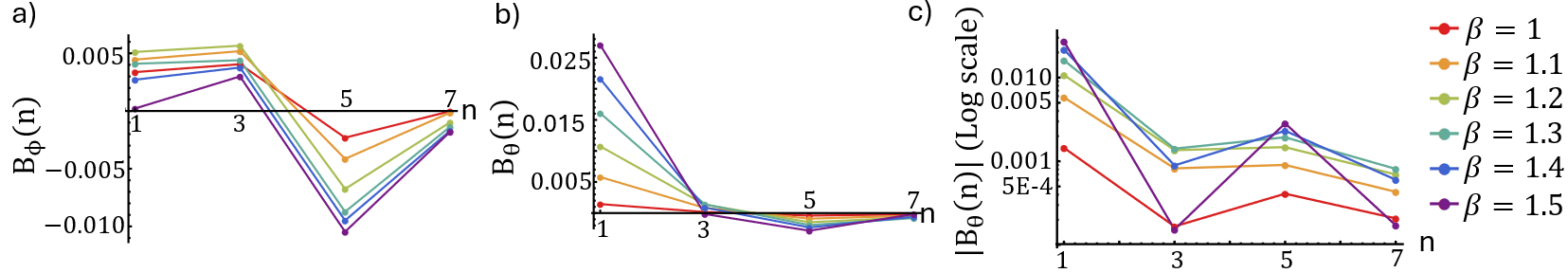}
    \caption{Dependence of the extracted Fourier coefficients on the SOC scale factor $\beta$ for equatorial magnetization ($\theta=\pi/2$). (a) Coefficients for $T_{\theta}(\phi)$, (b) coefficients for $T_{\phi}(\phi)$, and (c) $|T_{\phi}(\phi)|$ shown on a logarithmic scale to emphasize the nonlinear, harmonic-dependent SOC response.}
    \label{fig:socIncr}
\end{figure*}

Because SOC enters our Wannier Hamiltonian as an on-site term, we can explicitly tune its strength and examine how the torque harmonics respond. To provide a representative assessment (without recomputing the full $(\theta,\phi)$ torque map for each SOC value), we perform an SOC-scaling test on the equator $(\theta=\pi/2,\;0<\phi<2\pi)$, where the angular harmonics are most pronounced. Starting from the SOC parameters used in the main calculations (Pr: $0.11\,\mathrm{eV}$, Al: $0.02\,\mathrm{eV}$, Ge: $0.12\,\mathrm{eV}$), we scale all SOC terms by a global factor $\beta$ in the range $\beta=1.0$--$1.5$:

\[
\lambda_{\mathrm{Pr}}=\beta \times 0.11\,\mathrm{eV},\quad
\lambda_{\mathrm{Al}}=\beta \times 0.02\,\mathrm{eV},\quad
\lambda_{\mathrm{Ge}}=\beta \times 0.12\,\mathrm{eV}.
\]
We then extract Fourier harmonics of the equatorial torque components,
\[
T_{\theta}(\phi)=\sum_{n}\big[A_{\theta n}\sin(n\phi)+B_{\theta n}\cos(n\phi)\big],\]
\[
T_{\phi}(\phi)=\sum_{n}\big[A_{\phi n}\sin(n\phi)+B_{\phi n}\cos(n\phi)\big].
\]
Consistent with symmetry at $\theta=\pi/2$, only the cosine coefficients shown are nonzero in our geometry. Figure~\ref{fig:socIncr} summarizes the extracted coefficients $B_{\theta n}$ and $B_{\phi n}$ for several representative harmonics as a function of the SOC scale factor $\beta$. The main observation is that both lower-order and higher-order harmonic amplitudes vary with SOC strength, and the dependence is generally \emph{non-monotonic and harmonic-dependent}: different $n$ exhibit different trends with $\beta$, with crossings and sign/magnitude changes over $\beta=1.0$--$1.5$. For example, in Fig.~\ref{fig:socIncr}(b) the lower-order coefficients $B_{\phi 1}$ and $B_{\phi 3}$ are reduced in magnitude for $\beta=1.5$, whereas the higher-harmonic coefficient $B_{\phi 5}$ is enhanced. This non-monotonic behavior indicates that higher-harmonic torque components can be particularly sensitive to SOC-induced modifications of the electronic structure near the Fermi level.

\clearpage
\bibliography{references}{}

@PREAMBLE{
 "\providecommand{\noopsort}[1]{}"
 # "\providecommand{\singleletter}[1]{#1}%"
}

@article{Meng2019,
    author = {Meng, Biao and Wu, Hao and Qiu, Yang and Wang, Chunlei and Liu, Yong and Xia, Zhengcai and Yuan, Songliu and Chang, Haixin and Tian, Zhaoming},
    title = {Large anomalous Hall effect in ferromagnetic Weyl semimetal candidate PrAlGe},
    journal = {APL Materials},
    volume = {7},
    number = {5},
    pages = {051110},
    year = {2019},
    month = {05},
    issn = {2166-532X},
    doi = {10.1063/1.5090795},
    url = {https://doi.org/10.1063/1.5090795},
}

@article{Forslund2025,
  title = {Anomalous Hall Effect due to Magnetic Fluctuations in a Ferromagnetic Weyl Semimetal},
  author = {Forslund, Ola Kenji and Liu, Xiaoxiong and Shin, Soohyeon and Lin, Chun and Horio, Masafumi and Wang, Qisi and Kramer, Kevin and Mukherjee, Saumya and Kim, Timur and Cacho, Cephise and Wang, Chennan and Shang, Tian and Ukleev, Victor and White, Jonathan S. and Puphal, Pascal and Sassa, Yasmine and Pomjakushina, Ekaterina and Neupert, Titus and Chang, Johan},
  journal = {Phys. Rev. Lett.},
  volume = {134},
  issue = {12},
  pages = {126602},
  numpages = {8},
  year = {2025},
  month = {Mar},
  publisher = {American Physical Society},
  doi = {10.1103/PhysRevLett.134.126602},
  url = {https://link.aps.org/doi/10.1103/PhysRevLett.134.126602}
}

@article{Altermagnet_Review_PRX,
  title = {Emerging Research Landscape of Altermagnetism},
  author = {\ifmmode \check{S}\else \v{S}\fi{}mejkal, Libor and Sinova, Jairo and Jungwirth, Tomas},
  journal = {Phys. Rev. X},
  volume = {12},
  issue = {4},
  pages = {040501},
  numpages = {27},
  year = {2022},
  month = {Dec},
  publisher = {American Physical Society},
  doi = {10.1103/PhysRevX.12.040501},
  url = {https://link.aps.org/doi/10.1103/PhysRevX.12.040501}
}

@article{Anisotropy_exp2,
  title = {Critical behavior of the magnetic Weyl semimetal PrAlGe},
  author = {Liu, Wei and Zhao, Jun and Meng, Fanying and Rahman, Azizur and Qin, Yongliang and Fan, Jiyu and Pi, Li and Tian, Zhaoming and Du, Haifeng and Zhang, Lei and Zhang, Yuheng},
  journal = {Phys. Rev. B},
  volume = {103},
  issue = {21},
  pages = {214401},
  numpages = {8},
  year = {2021},
  month = {Jun},
  publisher = {American Physical Society},
  doi = {10.1103/PhysRevB.103.214401},
  url = {https://link.aps.org/doi/10.1103/PhysRevB.103.214401}
}

@article{Anisotropy_exp1,
  title = {Bulk single-crystal growth of the theoretically predicted magnetic Weyl semimetals $R\mathrm{AlGe}$ ($R$ = Pr, Ce)},
  author = {Puphal, Pascal and Mielke, Charles and Kumar, Neeraj and Soh, Y. and Shang, Tian and Medarde, Marisa and White, Jonathan S. and Pomjakushina, Ekaterina},
  journal = {Phys. Rev. Mater.},
  volume = {3},
  issue = {2},
  pages = {024204},
  numpages = {9},
  year = {2019},
  month = {Feb},
  publisher = {American Physical Society},
  doi = {10.1103/PhysRevMaterials.3.024204},
  url = {https://link.aps.org/doi/10.1103/PhysRevMaterials.3.024204}
}

@article{Kurita2020,
  title = {Systematic first-principles study of the on-site spin-orbit coupling in crystals},
  author = {Kurita, Kensuke and Koretsune, Takashi},
  journal = {Phys. Rev. B},
  volume = {102},
  issue = {4},
  pages = {045109},
  numpages = {7},
  year = {2020},
  month = {Jul},
  publisher = {American Physical Society},
  doi = {10.1103/PhysRevB.102.045109},
  url = {https://link.aps.org/doi/10.1103/PhysRevB.102.045109}
}

@article{
Tartaglia2020,
author = {Thomas A. Tartaglia  and Joseph N. Tang  and Jose L. Lado  and Faranak Bahrami  and Mykola Abramchuk  and Gregory T. McCandless  and Meaghan C. Doyle  and Kenneth S. Burch  and Ying Ran  and Julia Y. Chan  and Fazel Tafti },
title = {Accessing new magnetic regimes by tuning the ligand spin-orbit coupling in van der Waals magnets},
journal = {Science Advances},
volume = {6},
number = {30},
pages = {eabb9379},
year = {2020},
doi = {10.1126/sciadv.abb9379},
URL = {https://www.science.org/doi/abs/10.1126/sciadv.abb9379},
eprint = {https://www.science.org/doi/pdf/10.1126/sciadv.abb9379}}

@article{GGA+U,
  title = {Electron-energy-loss spectra and the structural stability of nickel oxide:  An LSDA+U study},
  author = {Dudarev, S. L. and Botton, G. A. and Savrasov, S. Y. and Humphreys, C. J. and Sutton, A. P.},
  journal = {Phys. Rev. B},
  volume = {57},
  issue = {3},
  pages = {1505--1509},
  numpages = {0},
  year = {1998},
  month = {Jan},
  publisher = {American Physical Society},
  doi = {10.1103/PhysRevB.57.1505},
  url = {https://link.aps.org/doi/10.1103/PhysRevB.57.1505}
}

@article{Fang_2025,
doi = {10.1088/1361-648X/adb192},
url = {https://dx.doi.org/10.1088/1361-648X/adb192},
year = {2025},
month = {feb},
publisher = {IOP Publishing},
volume = {37},
number = {15},
pages = {155801},
author = {Fang, Wuzhang and Schwartz, Edward and Kovalev, Alexey A and Belashchenko, K D},
title = {Spin–orbit torque in a three-fold-symmetric bilayer and its effect on magnetization dynamics},
journal = {Journal of Physics: Condensed Matter}
}

@article{Armitage2018RMP,
  title = {Weyl and Dirac semimetals in three-dimensional solids},
  author = {Armitage, N. P. and Mele, E. J. and Vishwanath, Ashvin},
  journal = {Rev. Mod. Phys.},
  volume = {90},
  issue = {1},
  pages = {015001},
  numpages = {57},
  year = {2018},
  month = {Jan},
  publisher = {American Physical Society},
  doi = {10.1103/RevModPhys.90.015001},
  url = {https://link.aps.org/doi/10.1103/RevModPhys.90.015001}
}

@article{NIELSEN1983,
title = {The Adler-Bell-Jackiw anomaly and Weyl fermions in a crystal},
journal = {Physics Letters B},
volume = {130},
number = {6},
pages = {389-396},
year = {1983},
issn = {0370-2693},
doi = {https://doi.org/10.1016/0370-2693(83)91529-0},
url = {https://www.sciencedirect.com/science/article/pii/0370269383915290},
author = {H.B. Nielsen and Masao Ninomiya}
}

@article{Mahfouzi2020,
  title = {Microscopic origin of spin-orbit torque in ferromagnetic heterostructures: A first-principles approach},
  author = {Mahfouzi, Farzad and Mishra, Rahul and Chang, Po-Hao and Yang, Hyunsoo and Kioussis, Nicholas},
  journal = {Phys. Rev. B},
  volume = {101},
  issue = {6},
  pages = {060405},
  numpages = {6},
  year = {2020},
  month = {Feb},
  publisher = {American Physical Society},
  doi = {10.1103/PhysRevB.101.060405},
  url = {https://link.aps.org/doi/10.1103/PhysRevB.101.060405}
}

@article{Sarkar2024,
author = {Shivam N. Kajale  and Thanh Nguyen  and Nguyen Tuan Hung  and Mingda Li  and Deblina Sarkar },
title = {Field-free deterministic switching of all–van der Waals spin-orbit torque system above room temperature},
journal = {Science Advances},
volume = {10},
number = {11},
pages = {eadk8669},
year = {2024},
doi = {10.1126/sciadv.adk8669},
URL = {https://www.science.org/doi/abs/10.1126/sciadv.adk8669},
eprint = {https://www.science.org/doi/pdf/10.1126/sciadv.adk8669}}

@article{Dieny2017,
  title = {Perpendicular magnetic anisotropy at transition metal/oxide interfaces and applications},
  author = {Dieny, B. and Chshiev, M.},
  journal = {Rev. Mod. Phys.},
  volume = {89},
  issue = {2},
  pages = {025008},
  numpages = {54},
  year = {2017},
  month = {Jun},
  publisher = {American Physical Society},
  doi = {10.1103/RevModPhys.89.025008},
  url = {https://link.aps.org/doi/10.1103/RevModPhys.89.025008}
}

@article{Xue2023,
  title = {Angular dependence of spin-orbit torque in monolayer ${\mathrm{Fe}}_{3}{\mathrm{GeTe}}_{2}$},
  author = {Xue, Fei and Stiles, Mark D. and Haney, Paul M.},
  journal = {Phys. Rev. B},
  volume = {108},
  issue = {14},
  pages = {144422},
  numpages = {12},
  year = {2023},
  month = {Oct},
  publisher = {American Physical Society},
  doi = {10.1103/PhysRevB.108.144422},
  url = {https://link.aps.org/doi/10.1103/PhysRevB.108.144422}
}

@article{Wang2022Cascadable,
author = {Lizheng Wang  and Junlin Xiong  and Bin Cheng  and Yudi Dai  and Fuyi Wang  and Chen Pan  and Tianjun Cao  and Xiaowei Liu  and Pengfei Wang  and Moyu Chen  and Shengnan Yan  and Zenglin Liu  and Jingjing Xiao  and Xianghan Xu  and Zhenlin Wang  and Youguo Shi  and Sang-Wook Cheong  and Haijun Zhang  and Shi-Jun Liang  and Feng Miao },
title = {Cascadable in-memory computing based on symmetric writing and readout},
journal = {Science Advances},
volume = {8},
number = {49},
pages = {eabq6833},
year = {2022},
doi = {10.1126/sciadv.abq6833},
URL = {https://www.science.org/doi/abs/10.1126/sciadv.abq6833},
eprint = {https://www.science.org/doi/pdf/10.1126/sciadv.abq6833}}

@article{Worledge2011,
author = {Worledge,D. C.  and Hu,G.  and Abraham,David W.  and Sun,J. Z.  and Trouilloud,P. L.  and Nowak,J.  and Brown,S.  and Gaidis,M. C.  and O’Sullivan,E. J.  and Robertazzi,R. P. },
title = {Spin torque switching of perpendicular Ta/CoFeB/MgO-based magnetic tunnel junctions},
journal = {Applied Physics Letters},
volume = {98},
number = {2},
pages = {022501},
year = {2011},
doi = {10.1063/1.3536482},

URL = { 
        https://doi.org/10.1063/1.3536482
    
},
eprint = { 
        https://doi.org/10.1063/1.3536482
    
}

}

@article{Shao2021,
   author = {Qiming Shao and Peng Li and Luqiao Liu and Hyunsoo Yang and Shunsuke Fukami and Armin Razavi and Hao Wu and Kang Wang and Frank Freimuth and Yuriy Mokrousov and Mark D. Stiles and Satoru Emori and Axel Hoffmann and Johan Akerman and Kaushik Roy and Jian-Ping Wang and See-Hun Yang and Kevin Garello and Wei Zhang},
   doi = {10.1109/TMAG.2021.3078583},
   issn = {0018-9464},
   issue = {7},
   journal = {IEEE Transactions on Magnetics},
   month = {7},
   pages = {1-39},
   title = {Roadmap of Spin–Orbit Torques},
   volume = {57},
   year = {2021},
}

@article{Grollier2020,
   author = {J. Grollier and D. Querlioz and K. Y. Camsari and K. Everschor-Sitte and S. Fukami and M. D. Stiles},
   doi = {10.1038/s41928-019-0360-9},
   issn = {2520-1131},
   issue = {7},
   journal = {Nature Electronics},
   month = {7},
   pages = {360-370},
   title = {Neuromorphic spintronics},
   volume = {3},
   year = {2020},
}

@article{Hoffmann2022,
author = {Hoffmann,Axel  and Ramanathan,Shriram  and Grollier,Julie  and Kent,Andrew D.  and Rozenberg,Marcelo J.  and Schuller,Ivan K.  and Shpyrko,Oleg G.  and Dynes,Robert C.  and Fainman,Yeshaiahu  and Frano,Alex  and Fullerton,Eric E.  and Galli,Giulia  and Lomakin,Vitaliy  and Ong,Shyue Ping  and Petford-Long,Amanda K.  and Schuller,Jonathan A.  and Stiles,Mark D.  and Takamura,Yayoi  and Zhu,Yimei },
title = {Quantum materials for energy-efficient neuromorphic computing: Opportunities and challenges},
journal = {APL Materials},
volume = {10},
number = {7},
pages = {070904},
year = {2022},
doi = {10.1063/5.0094205},
URL = { 
        https://doi.org/10.1063/5.0094205
},
}

@article{Kao2022,
   author = {I-Hsuan Kao and Ryan Muzzio and Hantao Zhang and Menglin Zhu and Jacob Gobbo and Sean Yuan and Daniel Weber and Rahul Rao and Jiahan Li and James H. Edgar and Joshua E. Goldberger and Jiaqiang Yan and David G. Mandrus and Jinwoo Hwang and Ran Cheng and Jyoti Katoch and Simranjeet Singh},
   doi = {10.1038/s41563-022-01275-5},
   issn = {1476-1122},
   issue = {9},
   journal = {Nature Materials},
   month = {9},
   pages = {1029-1034},
   title = {Deterministic switching of a perpendicularly polarized magnet using unconventional spin–orbit torques in WTe2},
   volume = {21},
   year = {2022},
}

@article{Xue2021SOT_AFM,
  title = {Intrinsic staggered spin-orbit torque for the electrical control of antiferromagnets: Application to ${\mathrm{CrI}}_{3}$},
  author = {Xue, Fei and Haney, Paul M.},
  journal = {Phys. Rev. B},
  volume = {104},
  issue = {22},
  pages = {224414},
  numpages = {12},
  year = {2021},
  month = {Dec},
  publisher = {American Physical Society},
  doi = {10.1103/PhysRevB.104.224414},
  url = {https://link.aps.org/doi/10.1103/PhysRevB.104.224414}
}

@article{Brataas2019,
  title = {Current Control of Magnetism in Two-Dimensional ${\mathrm{Fe}}_{3}{\mathrm{GeTe}}_{2}$},
  author = {Johansen, \O{}yvind and Risingg\aa{}rd, Vetle and Sudb\o{}, Asle and Linder, Jacob and Brataas, Arne},
  journal = {Phys. Rev. Lett.},
  volume = {122},
  issue = {21},
  pages = {217203},
  numpages = {6},
  year = {2019},
  month = {May},
  publisher = {American Physical Society},
  doi = {10.1103/PhysRevLett.122.217203},
  url = {https://link.aps.org/doi/10.1103/PhysRevLett.122.217203}
}

@article{Belashchenko2020,
  title = {Interfacial contributions to spin-orbit torque and magnetoresistance in ferromagnet/heavy-metal bilayers},
  author = {Belashchenko, K. D. and Kovalev, Alexey A. and van Schilfgaarde, M.},
  journal = {Phys. Rev. B},
  volume = {101},
  issue = {2},
  pages = {020407},
  numpages = {6},
  year = {2020},
  month = {Jan},
  publisher = {American Physical Society},
  doi = {10.1103/PhysRevB.101.020407},
  url = {https://link.aps.org/doi/10.1103/PhysRevB.101.020407}
}

@article{SLONCZEWSKI1996,
title = "Current-driven excitation of magnetic multilayers",
journal = "Journal of Magnetism and Magnetic Materials",
volume = "159",
number = "1",
pages = "L1 - L7",
year = "1996",
issn = "0304-8853",
doi = "https://doi.org/10.1016/0304-8853(96)00062-5",
url = "http://www.sciencedirect.com/science/article/pii/0304885396000625",
author = "J.C. Slonczewski"
}

@article{Manchon2019review,
  title = {Current-induced spin-orbit torques in ferromagnetic and antiferromagnetic systems},
  author = {Manchon, A. and \ifmmode \check{Z}\else \v{Z}\fi{}elezn\'y, J. and Miron, I. M. and Jungwirth, T. and Sinova, J. and Thiaville, A. and Garello, K. and Gambardella, P.},
  journal = {Rev. Mod. Phys.},
  volume = {91},
  issue = {3},
  pages = {035004},
  numpages = {80},
  year = {2019},
  month = {Sep},
  publisher = {American Physical Society},
  doi = {10.1103/RevModPhys.91.035004},
  url = {https://link.aps.org/doi/10.1103/RevModPhys.91.035004}
}

@article{Xue2020SOT,
  title = {Unconventional spin-orbit torque in transition metal dichalcogenide--ferromagnet bilayers from first-principles calculations},
  author = {Xue, Fei and Rohmann, Christoph and Li, Junwen and Amin, Vivek and Haney, Paul},
  journal = {Phys. Rev. B},
  volume = {102},
  issue = {1},
  pages = {014401},
  numpages = {10},
  year = {2020},
  month = {Jul},
  publisher = {American Physical Society},
  doi = {10.1103/PhysRevB.102.014401},
  url = {https://link.aps.org/doi/10.1103/PhysRevB.102.014401}
}

@article{go2020theory,
   title = {Theory of current-induced angular momentum transfer dynamics in spin-orbit coupled systems},
  author = {Go, Dongwook and Freimuth, Frank and Hanke, Jan-Philipp and Xue, Fei and Gomonay, Olena and Lee, Kyung-Jin and Bl\"ugel, Stefan and Haney, Paul M. and Lee, Hyun-Woo and Mokrousov, Yuriy},
  journal = {Phys. Rev. Res.},
  volume = {2},
  issue = {3},
  pages = {033401},
  numpages = {24},
  year = {2020},
  month = {Sep},
  publisher = {American Physical Society},
  doi = {10.1103/PhysRevResearch.2.033401},
  url = {https://link.aps.org/doi/10.1103/PhysRevResearch.2.033401}
}

@Article{Garello2013,
author={Garello, Kevin
and Miron, Ioan Mihai
and Avci, Can Onur
and Freimuth, Frank
and Mokrousov, Yuriy
and Bl{\"u}gel, Stefan
and Auffret, St{\'e}phane
and Boulle, Olivier
and Gaudin, Gilles
and Gambardella, Pietro},
title={Symmetry and magnitude of spin-orbit torques in ferromagnetic heterostructures},
journal={Nature Nanotechnology},
year={2013},
volume={8},
number={8},
pages={587-593},
issn={1748-3395},
doi={10.1038/nnano.2013.145},
url={https://doi.org/10.1038/nnano.2013.145}
}

@article{miron2011perpendicular,
  title={Perpendicular switching of a single ferromagnetic layer induced by in-plane current injection},
  author={Miron, Ioan Mihai and Garello, Kevin and Gaudin, Gilles and Zermatten, Pierre-Jean and Costache, Marius V and Auffret, St{\'e}phane and Bandiera, S{\'e}bastien and Rodmacq, Bernard and Schuhl, Alain and Gambardella, Pietro},
  journal={Nature},
  volume={476},
  number={7359},
  pages={189--193},
  year={2011},
  publisher={Nature Publishing Group},
  doi={https://doi.org/10.1038/nature10309},
}

@article{Freimuth2014,
  title = {Spin-orbit torques in Co/Pt(111) and Mn/W(001) magnetic bilayers from first principles},
  author = {Freimuth, Frank and Bl\"ugel, Stefan and Mokrousov, Yuriy},
  journal = {Phys. Rev. B},
  volume = {90},
  issue = {17},
  pages = {174423},
  numpages = {10},
  year = {2014},
  month = {Nov},
  publisher = {American Physical Society},
  doi = {10.1103/PhysRevB.90.174423},
  url = {https://link.aps.org/doi/10.1103/PhysRevB.90.174423}
}

@article{Mahfouzi2018,
  title = {First-principles study of the angular dependence of the spin-orbit torque in Pt/Co and Pd/Co bilayers},
  author = {Mahfouzi, Farzad and Kioussis, Nicholas},
  journal = {Phys. Rev. B},
  volume = {97},
  issue = {22},
  pages = {224426},
  numpages = {7},
  year = {2018},
  month = {Jun},
  publisher = {American Physical Society},
  doi = {10.1103/PhysRevB.97.224426},
  url = {https://link.aps.org/doi/10.1103/PhysRevB.97.224426}
}

@Article{MacNeill2016,
author={MacNeill, D.
and Stiehl, G. M.
and Guimaraes, M. H. D.
and Buhrman, R. A.
and Park, J.
and Ralph, D. C.},
title={Control of spin-orbit torques through crystal symmetry in WTe2/ferromagnet bilayers},
journal={Nature Physics},
year={2016},
month={Nov},
day={07},
publisher={Nature Publishing Group},
volume={13},
pages={300-305},
url={https://doi.org/10.1038/nphys3933}
}

@article{Wannier90,
title = "An updated version of wannier90: A tool for obtaining maximally-localised Wannier functions",
journal = "Computer Physics Communications",
volume = "185",
number = "8",
pages = "2309 - 2310",
year = "2014",
issn = "0010-4655",
doi = "https://doi.org/10.1016/j.cpc.2014.05.003",
url = "http://www.sciencedirect.com/science/article/pii/S001046551400157X",
author = "Arash A. Mostofi and Jonathan R. Yates and Giovanni Pizzi and Young-Su Lee and Ivo Souza and David Vanderbilt and Nicola Marzari"
}

@article{Belashchenko2019,
  title = {First-principles calculation of spin-orbit torque in a Co/Pt bilayer},
  author = {Belashchenko, K. D. and Kovalev, Alexey A. and van Schilfgaarde, M.},
  journal = {Phys. Rev. Materials},
  volume = {3},
  issue = {1},
  pages = {011401},
  numpages = {6},
  year = {2019},
  month = {Jan},
  publisher = {American Physical Society},
  doi = {10.1103/PhysRevMaterials.3.011401},
  url = {https://link.aps.org/doi/10.1103/PhysRevMaterials.3.011401}
}

@article{liu2012current,
   title = {Current-Induced Switching of Perpendicularly Magnetized Magnetic Layers Using Spin Torque from the Spin Hall Effect},
  author = {Liu, Luqiao and Lee, O. J. and Gudmundsen, T. J. and Ralph, D. C. and Buhrman, R. A.},
  journal = {Phys. Rev. Lett.},
  volume = {109},
  issue = {9},
  pages = {096602},
  numpages = {5},
  year = {2012},
  month = {Aug},
  publisher = {American Physical Society},
  doi = {10.1103/PhysRevLett.109.096602},
  url = {https://link.aps.org/doi/10.1103/PhysRevLett.109.096602}
}

@article{liu2012spin,
author = {Luqiao Liu  and Chi-Feng Pai  and Y. Li  and H. W. Tseng  and D. C. Ralph  and R. A. Buhrman },
title = {Spin-Torque Switching with the Giant Spin Hall Effect of Tantalum},
journal = {Science},
volume = {336},
number = {6081},
pages = {555-558},
year = {2012},
doi = {10.1126/science.1218197},
}

@article{VASP,
  title = {Efficient iterative schemes for ab initio total-energy calculations using a plane-wave basis set},
  author = {Kresse, G. and Furthm\"uller, J.},
  journal = {Phys. Rev. B},
  volume = {54},
  issue = {16},
  pages = {11169--11186},
  numpages = {0},
  year = {1996},
  month = {Oct},
  publisher = {American Physical Society},
  doi = {10.1103/PhysRevB.54.11169},
  url = {https://link.aps.org/doi/10.1103/PhysRevB.54.11169}
}

@article{Chang2018,
  author = {Chang, G. and Singh, B. and Xu, S.-Y. and Bian, G. and Huang, S.-M. and Hsu, C.-H. and Belopolski, I. and Alidoust, N. and Sanchez, D. S. and Zheng, H. and Lu, H. and Zhang, X. and Bian, Y. and Chang, T.-R. and Jeng, H.-T. and Bansil, A. and Hsu, H. and Jia, S. and Neupert, T. and Hasan, M. Z.},
  title = {Magnetic and noncentrosymmetric Weyl fermion semimetals in the},
  journal = {Physical Review B},
  volume = {97},
  number = {4},
  year = {2018},
  doi = {10.1103/physrevb.97.041104}
}

@article{Sanchez2020,
  author = {Sanchez, D. S. and Chang, G. and Belopolski, I. and Lu, H. and Yin, J.-X. and Alidoust, N. and Xu, X. and Cochran, T. A. and Zhang, X. and Bian, Y. and Zhang, S. S. and Liu, Y.-Y. and Ma, J. and Bian, G. and Lin, H. and Xu, S.-Y. and Jia, S. and Hasan, M. Z.},
  title = {Observation of Weyl fermions in a magnetic non-centrosymmetric crystal},
  journal = {Nature Communications},
  volume = {11},
  number = {1},
  year = {2020},
  doi = {10.1038/s41467-020-16879-1}
}

@article{WU2017,
title = "WannierTools : An open-source software package for novel topological materials",
journal = "Computer Physics Communications",
volume = "224",
pages = "405 - 416",
year = "2018",
issn = "0010-4655",
doi = "https://doi.org/10.1016/j.cpc.2017.09.033",
url = "http://www.sciencedirect.com/science/article/pii/S0010465517303442",
author = "QuanSheng Wu and ShengNan Zhang and Hai-Feng Song and Matthias Troyer and Alexey A. Soluyanov"
}

@article{Skomski2009,
  author       = {Skomski, R. and Sellmyer, D. J.},
  title        = {Anisotropy of rare-earth magnets},
  journal      = {Journal of Rare Earths},
  year         = {2009},
  volume       = {27},
  number       = {4},
  pages        = {675--679},
  doi          = {10.1016/S1002-0721(08)60314-2},
}

@article{Blochl1994,
  author       = {Bl{\"o}chl, P. E.},
  title        = {Projector augmented-wave method},
  journal      = {Physical Review B},
  year         = {1994},
  volume       = {50},
  number       = {24},
  pages        = {17953--17979},
  doi          = {10.1103/PhysRevB.50.17953},
  publisher    = {American Physical Society}
}

@article{Kresse1999,
  author       = {Kresse, G. and Joubert, D.},
  title        = {From ultrasoft pseudopotentials to the projector augmented-wave method},
  journal      = {Physical Review B},
  year         = {1999},
  volume       = {59},
  number       = {3},
  pages        = {1758--1775},
  doi          = {10.1103/PhysRevB.59.1758},
  publisher    = {American Physical Society}
}

@article{Destraz2020,
  author    = {Destraz, D. and Das, L. and Tsirkin, S. S. and Xu, Y. and Neupert, T. and Chang, J. and Schilling, A. and Grushin, A. G. and Kohlbrecher, J. and Keller, L. and Puphal, P. and Pomjakushina, E. and White, J. S.},
  title     = {Magnetism and anomalous transport in the Weyl semimetal PrAlGe: Possible route to axial gauge fields},
  journal   = {npj Quantum Materials},
  volume    = {5},
  number    = {1},
  pages     = {5},
  year      = {2020},
  publisher = {Springer Nature},
  doi       = {10.1038/s41535-019-0207-7},
  url       = {https://doi.org/10.1038/s41535-019-0207-7}
}

@article{Yang2022,
  author       = {Yang, R. and Corasaniti, M. and Le, C. C. and Yue, C. and Hu, Z. and Hu, J. P. and Petrovic, C. and Degiorgi, L.},
  title        = {Charge dynamics of a noncentrosymmetric magnetic Weyl semimetal},
  journal      = {npj Quantum Materials},
  year         = {2022},
  volume       = {7},
  number       = {1},
  pages        = {5},
  doi          = {10.1038/s41535-022-00507-w},
  publisher    = {Springer Nature}
}

@article{Liu2021,
  author       = {Liu, W. and Zhao, J. and Meng, F. and Rahman, A. and Qin, Y. and Fan, J. and Pi, L. and Tian, Z. and Du, H. and Zhang, L. and Zhang, Y.},
  title        = {Critical behavior of the magnetic Weyl semimetal PrAlGe},
  journal      = {Physical Review B},
  year         = {2021},
  volume       = {103},
  number       = {21},
  pages        = {214401},
  doi          = {10.1103/PhysRevB.103.214401},
  publisher    = {American Physical Society}
}

@article{Zhi2022WannSymm,
  author       = {Zhi, G.-X. and Xu, C. and Wu, S.-Q. and Ning, F. and Cao, C.},
  title        = {WannSymm: A symmetry analysis code for Wannier orbitals},
  journal      = {Computer Physics Communications},
  volume       = {271},
  pages        = {108196},
  year         = {2022},
  doi          = {10.1016/j.cpc.2021.108196}
}
\end{document}